\documentclass[reqno,11pt]{amsart}
\usepackage{amsmath, latexsym, amsfonts, amssymb, amsthm, amscd}
\usepackage{graphics,epsf,psfrag}
\numberwithin{equation}{section}

\newtheorem{theorem}{Theorem}[section]

\newtheorem{proposition}[theorem]{Proposition}

\newtheorem{rem}[theorem]{Remark}

\newcommand{\ind}{\mathbf{1}}

\newcommand{\R}{\mathbb{R}}
\newcommand{\Z}{\mathbb{Z}}
\newcommand{\N}{\mathbb{N}}
\renewcommand{\tilde}{\widetilde}
\renewcommand{\hat}{\widehat}

\newcommand{\cH}{{\ensuremath{\mathcal H}} }

\newcommand{\cN}{{\ensuremath{\mathcal N}} }
\newcommand{\cL}{{\ensuremath{\mathcal L}} }

\newcommand{\cD}{{\ensuremath{\mathcal D}} }

\newcommand{\bP}{{\ensuremath{\mathbf P}} }
\newcommand{\bE}{{\ensuremath{\mathbf E}} }

\DeclareMathSymbol{\leqslant}{\mathalpha}{AMSa}{"36} 
\DeclareMathSymbol{\geqslant}{\mathalpha}{AMSa}{"3E} 
\DeclareMathSymbol{\eset}{\mathalpha}{AMSb}{"3F}     
\newcommand{\dd}{\,\text{\rm d}}             
\newcommand{\sumtwo}[2]{\sum_{\substack{#1 \\ #2}}} 

\newcommand{\bbE}{{\ensuremath{\mathbb E}} }

\newcommand{\bbP}{{\ensuremath{\mathbb P}} }

\newcommand{\ga}{\alpha}
\newcommand{\gb}{\beta}
\newcommand{\gd}{\delta}
\newcommand{\gep}{\varepsilon}       

\newcommand{\go}{\omega}

\makeatletter
\def\captionfont@{\footnotesize}
\def\captionheadfont@{\scshape}

\long\def\@makecaption#1#2{%
  \vspace{2mm}
  \setbox\@tempboxa\vbox{\color@setgroup
    \advance\hsize-6pc\noindent
    \captionfont@\captionheadfont@#1\@xp\@ifnotempty\@xp
        {\@cdr#2\@nil}{.\captionfont@\upshape\enspace#2}%
    \unskip\kern-6pc\par
    \global\setbox\@ne\lastbox\color@endgroup}%
  \ifhbox\@ne 
    \setbox\@ne\hbox{\unhbox\@ne\unskip\unskip\unpenalty\unkern}%
  \fi
  \ifdim\wd\@tempboxa=\z@ 
    \setbox\@ne\hbox to\columnwidth{\hss\kern-6pc\box\@ne\hss}%
  \else 
    \setbox\@ne\vbox{\unvbox\@tempboxa\parskip\z@skip
        \noindent\unhbox\@ne\advance\hsize-6pc\par}%
\fi
  \ifnum\@tempcnta<64 
    \addvspace\abovecaptionskip
    \moveright 3pc\box\@ne
  \else 
    \moveright 3pc\box\@ne
    \nobreak
    \vskip\belowcaptionskip
  \fi
\relax
}
\makeatother
\def\writefig#1 #2 #3 {\rlap{\kern #1 truecm
\raise #2 truecm \hbox{#3}}}

\newcommand{\tf}{\textsc{f}}

\newcommand{\tc}{\textsc{c}}

\newcommand{\rc}{\mathtt c}
\newcommand{\rf}{\mathtt f}

\newcommand{\Kbar}{{\overline{K}}}

\begin{document}

\title[Disordered pinning models]{
Renewal sequences, disordered potentials,
\\ and pinning phenomena
}

\author{Giambattista Giacomin}
\address{
  Universit{\'e} Paris  Diderot (Paris 7) and Laboratoire de Probabilit{\'e}s et Mod\`eles Al\'eatoires (CNRS U.M.R. 7599),
U.F.R.        de        Math\'ematiques, Case 7012 (site Chevaleret),
                75205 Paris cedex 13, France
                }
\email{giacomin\@@math.jussieu.fr}

\date{\today}

\begin{abstract} 
We give an overview of the state of the art of the analysis of 
 disordered  models of pinning on a defect line. This class of models 
includes a number of well known and much studied systems
(like polymer pinning on a defect line, wetting of interfaces 
on a disordered substrate and the Poland-Scheraga model 
of DNA denaturation). A remarkable aspect is that, in 
absence of disorder, all the models in this class 
are exactly solvable 
and they display a localization-delocalization transition that one 
understands in full detail. 
Moreover the behavior of such systems
near criticality is controlled by a parameter and 
one observes, by
tuning the parameter,
 the {\sl full spectrum} of critical behaviors, ranging
from first order to infinite order transitions.    
This is therefore an ideal set-up in which to address the 
question of the effect of disorder on the phase transition,
notably on  critical properties. We will review  recent results
that show
that the physical prediction that goes under the name
of {\sl Harris criterion} is indeed fully correct for pinning models. 
Beyond summarizing the results, we will sketch most 
of the arguments of proof. 
\end{abstract}

\maketitle

\tableofcontents

\section{Pinning and disorder: models and motivations}
\subsection{The basic example: pinning of simple random walks}
It is somewhat customary to introduce pinning  models by 
talking of pinning of simple random walks (SRW). This is due to a
number of reasons, like the widespread grasp on SRW, or 
 the fact 
 that modeling several pinning phenomena naturally leads to random walk pinning, as we will see. 
 However, we will see also that, in a sense, the SRW case
 is the hardest to deal with: nonetheless, 
 we are going to follow the tradition and start from  SRW pinning.
 
 Let $S:=\{S_n\}_{n=0,1, \ldots}$ be a sequence of random variables
 such that $S_0=0$ and such that $\{ S_{n}- S_{n-1}\}_{n=1, 2 , \ldots }$ are IID
 ({\sl i.e.}, independent and identically distributed) symmetric random variables
taking only  the values $+1$ and $-1$.
The {\sl disorder} is given by a  sequence  $\go:=\{\go_n\}
_{n=1,2, \ldots}$ of real numbers and we will play with two
real parameters $\gb$ and $h$.
We actually assume that $\go$ is a realization of an IID sequence of
standard Gaussian variables  
(see Remark~\ref{rem:nonGauss} for some comments on generalizations). 
We call $\bbP$ the law of $\go$ and we denote by $\theta$
the left-shift operator on $\R^\N$: $(\theta \go)_n= \go_{n+1}$. 
Our aim is to study the probability measure
$P_{N, \go}$ ($N$ is a positive integer: we will be  interested
in the limit $N\to \infty$)  defined as
\begin{equation}
\label{eq:basic0}
P_{N, \go} (s_0,s_1, \ldots, s_N)\, :=\, \frac 1{Z_{N, \go}}
\exp\left(\sum_{n=1}^N \left( \gb \go_n +h \right) \ind_{s_n=0}\right)
P (s_0,s_1,  \ldots, s_N),
\end{equation}
where 
\begin{enumerate}
\item
$P (s_1, s_2, \ldots, s_N)=(1/2)^N$
if and only if $s_0=0$ and $\vert s_{n}-s_{n-1}\vert=1$ for
$n=1, 2, \ldots, N$;
\item $Z_{N, \go}$ is the normalization constant ({\sl partition function}), that is
\begin{equation}
Z_{N, \go}\, =\, \sum_{s_0,s_1,  \ldots, s_N}
\exp\left(\sum_{n=1}^N \left( \gb \go_n +h \right) \ind_{s_n=0}\right)
P (s_0,s_1,  \ldots, s_N).
\end{equation}
\end{enumerate}

\begin{figure}[htp]
\begin{center}
\leavevmode
\epsfxsize =13 cm
\psfragscanon
\psfrag{0}[c]{$0$}
\psfrag{n}[c]{$n$}
\psfrag{Sn}[c]{$S_n$}
\psfrag{go1}[c]{\tiny $\gb\go_2+h$}
\psfrag{go2}[c]{\tiny $\gb\go_4+h$}
\psfrag{go3}[c]{\tiny $\gb\go_6+h$}
\psfrag{A}[c]{\huge \tt A}
\psfrag{B}[c]{\huge \tt B}
\epsfbox{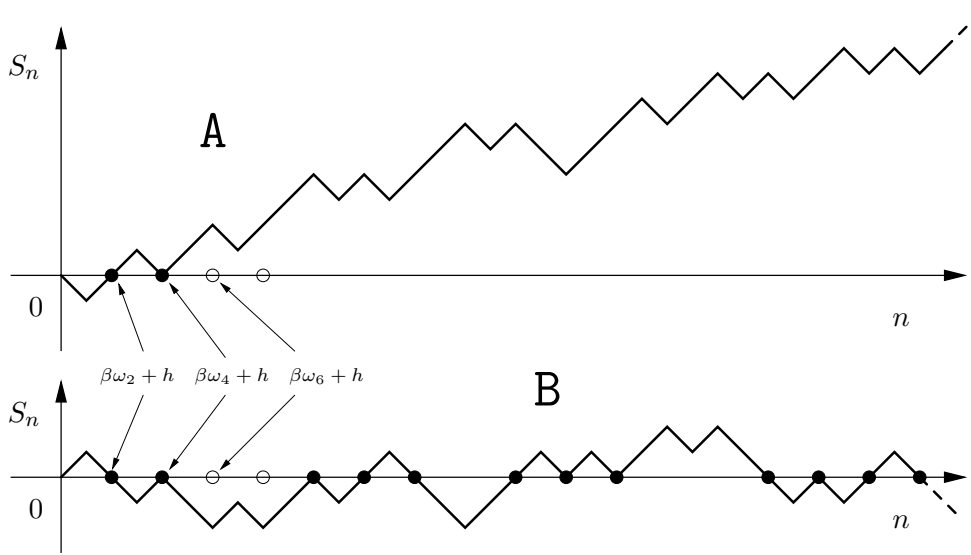}
\end{center}
\caption{\label{fig:SRW} 
Two trajectories sampled from $\bP_{N,\go}$ for different values of $\gb$ and  $h$, with  $N$ {\sl very large},
and represented as directed polymers, in the sense that we plot the function $n \mapsto S_n$ or, equivalently, we
look at the walk $\{ (n, S_n)\}_{n=0,1, \ldots}$ with one deterministic component. 
The case {\tt A} sketches a delocalized trajectory that is what one observes when $h$ is, for example,
 negative and large (one should first think of the homogeneous case $\gb=0$). Since the SRW is a periodic
 Markov chain, the origin is visited only at even times, so only $\go_2, \go_4, \go_6, \ldots$ 
 play a role and they are the only {\sl charges} (a charge at $n $ is the quantity 
 $\gb \go_n +h$) marked in the drawing, with a filled circle if they are visited and with an empty
 circle if they are not visited: not all the unvisited charges are marked). The distinctive aspect of
 case {\tt A}  is that there are just a few {\sl contacts}, {\sl i.e.} visits to the origin, and then the walk
 resembles a walk conditioned not to hit zero. 
 The case {\tt B} is instead what one observes when $h$ is positive and large: the number of contacts
 is large, as a matter of fact the drawing wants to suggest that there is 
 a positive
 density
 of contacts. It is natural to call such a regime {\sl localized},
 in contrast to the previous one that  we call {\sl delocalized}. 
 Note that both   case {\tt A} and {\tt B} are atypical for the free walk ($\gb=h=0$)
 in which there is a zero density of contacts but they are spread through the system and the walk 
 certainly does not stay on one side of the axis as in case {\tt A}. It is important to remark 
 that we have a full up--down symmetry and this implies that in the delocalized regime  {\tt A}
 the walk is either delocalized above or below the axis with probability $1/2$. Of course at this
 stage it is highly unclear that one observes either localized or delocalized trajectories (for
 typical $\go$) and to a certain extent this is not correct because one has to exclude the
 so called {\sl critical regime}, which however appears only at exceptional values of $\gb $ and $h$
 (the {\sl phase transition} point, or {\sl critical} point). We have of course avoided
 the delicate issue of the role of disorder, at the hearth of this presentation. Here we will simply content
 ourselves with pointing out that, for example, even when $h$ is very large and negative (pushing
 thus toward delocalization) any amount of disorder, {\sl i.e.} $\gb>0$,
 yields a positive density of sites in which $\gb\go_n +h$ is positive and therefore attractive.
 There could therefore be a {\sl smart targeting strategy} of the polymer in placing the contacts at these sites, leading thus to
 localization.   
}
\end{figure}
  
 We will actually prefer a slightly different definition of the model, namely
given the sequence $s=\{s_0,s_1, \ldots\}$ we set
\begin{equation}
\label{eq:basic}
\frac{\dd \bP_{N, \go}}{\dd\bP} (s) \, :=\, \frac 1{Z_{N, \go}}
\exp\left(\sum_{n=1}^N \left( \gb \go_n +h \right) \ind_{s_n=0}\right).
\end{equation}
Notice that this time $\bP_{N, \go}$ is a measure on (infinite) sequences,
namely the trajectory of the walk all the way to infinity, while in  
  \eqref{eq:basic0} we had defined a measure only up to {\sl step} 
  (or {\sl time}) $N$.
 As a matter of fact, if we consider {\sl cylindrical} events of the type
 $E=\{s=\{s_n\}_{n=0,1, \ldots}:\, s_0=t_0, s_1=t_1, \ldots, s_N=t_N\}$,
 then the measure of $E$ under $\bP_{N, \go}$ defined in \eqref{eq:basic}
 coincides with $P_{N, \go} (t_0,t_1, \ldots, t_N)$.
 
 \smallskip
 \begin{rem}
 \label{rem:quenched} \rm
 It is worth stressing that, unless $\gb=0$, in this model there are two sources
 of randomness: the polymer chain is modeled by a random walk with law $\bP$ and the
 disorder is a typical realization of the random sequence $\go$ with law $\bbP$.
 These two sources of randomness are treated in very different ways: $\go$ is {\sl quenched},
 that is chosen once and for all, while the polymer location fluctuates and in fact we study
 the distribution of $S$ under $\bP_{N ,  \go}$. 
 \end{rem}
 \smallskip
 
 As it is well known, the Markov process $S$ is null-recurrent, namely every
 site of the state space $\Z$ is visited (infinitely often) $\bP$-almost surely,
 but the expectation of the time between successive visits to a given
  site is 
 infinite. Let us be more explicit about this last concept and let us introduce, for 
 $m\in \Z$,
 the random variable $\tau_1(m):= \inf \{ n>0:\, S_n=m\}$ and, for $j>1$, 
 also $\tau_j(m):= \inf \{ n>\tau_{j-1}(m):\, S_n=m\}$.
 It is then a direct consequence of the (strong) Markov property that
 $\{\tau_{j+1}(m)-\tau_{j}(m)\}_{j=1,2, \ldots}$ is a sequence of IID  random variables. It should be also clear
 that the law of 
 $\{\tau_{j+1}(m)-\tau_{j}(m)\}_{j=1,2, \ldots}$ 
 does not depend on the value of $m$:
  we are going to denote $\tau_j(0)$ simply by $\tau_j$ and, since $S_0=0$,
  we are setting $\tau_0=0$. The recurrent character of $S$
  boils simply down to the fact that $\sum_{n}\bP(\tau_1=n)=1$ and
   the recurrence is of null type  because $\bE[\tau_1]=+\infty$: the distribution
   of $\tau_1$ is known in detail \cite[Ch.~III]{cf:Feller1} and in particular
   \begin{equation}
 \label{eq:tauSRW}
 \bP( \tau_1=2n) \stackrel{n\to \infty}\sim \frac{c}{n^{3/2}}, 
   \end{equation}
   where $c=1/\sqrt{4\pi}$ and we have introduced the notation $a_n \stackrel{n\to \infty}\sim  b_n$ for
   $\lim_{n \to \infty} a_n/b_n =1$. Since (clearly) $\bE[\tau_1]=\infty$, the classical 
   Renewal  Theorem (see \eqref{eq:RT} below) 
   tells us that the expected number of visits to $0$ of $S$
   up to time $N$ is $o(N)$ (a more precise analysis shows that it is
   of the order of $\sqrt{N}$ , see  \cite[Ch.~III]{cf:Feller1}
   or Theorem~\ref{th:renewal} below).

  As we shall see, 
the trajectories of the process $S$ are very strongly affected 
if $\gb$ or $h$ are non zero and, except for {\sl critical cases}, 
what happens is roughly that, in the limit as $N\to \infty$,
under $\bP_{N , \go}$ the expected number of the visits paid by $S$ to 
 $0$ is of the order of $N$, or it is much smaller than $\sqrt{N}$ 
 (in some cases one can show that they are $O(1)$). 
  A first glimpse at these different scenarios can be found in Figure~\ref{fig:SRW}.

\subsection{The general model: renewal pinning}
We have introduced  the $\tau$ sequence in the previous subsection 
in order to give some intuition about the model, but its interest goes well beyond.
A look at
\eqref{eq:basic} suffices to realize that $Z_{N , \go}$ can be expressed simply in terms 
of $\tau$:
\begin{equation}
Z_{N , \go}\, =\, \bE \exp\left( \sum_{n=1}^N (\gb \go_n +h) \ind_{n \in \tau}\right),
\end{equation} 
where we have introduced a notation that comes from looking
at $\tau= \{ \tau_j\}_{j=0,1, \ldots} $ as  a random subset of $\N \cup \{0\}$,
so that $n \in \tau$ means that there exists $j$ such that $\tau_j=n$.
Therefore the model in \eqref{eq:basic} is just a particular case of when
$\tau$ is a general discrete renewal:
\begin{equation}
\label{eq:basicf}
\frac{\dd \bP_{N, \go}^\rf}{\dd\bP} (\tau) \, :=\, \frac 1{Z_{N, \go}^\rf}
\exp\left(\sum_{n=1}^N \left( \gb \go_n +h \right) \ind_{n \in \tau}\right),
\end{equation}
where 
  the superscript $\rf$, that stands for {\sl free}, has been introduced because 
a slightly different version of the model is going to be relevant too:
\begin{equation}
\label{eq:basicc}
\frac{\dd \bP_{N, \go}^\rc}{\dd\bP} (\tau) \, :=\, \frac 1{Z_{N, \go}^\rc}
\exp\left(\sum_{n=1}^N \left( \gb \go_n +h \right) \ind_{n \in \tau}\right)
\ind_{N\in \tau},
\end{equation}
and $\rc$ stands for {\sl constrained}.
 Let us stress that by (discrete)
renewal process $\tau$ we simply mean 
a sequence of random variables with (positive and integer valued)
IID increments: we call these increments {\sl inter-arrival} variables. 

We will not be interested in the most general discrete renewal, but we will rather focus on the case
in which
\begin{equation}
\label{eq:K}
K(n)\, :=\, \bP\left( \tau_1=n\right)\, =\, \frac{L(n)}{n^{1+\ga}}, \ \ \  n\in \N =\{1,2, \ldots\}
\end{equation}
where $\ga$ is a positive number and 
\begin{equation}
\label{eq:L}
\lim_{n \to \infty} L(n)\, =\, \tc_K \,>\, 0.
\end{equation}
We call $K(\cdot)$ {\sl inter-arrival distribution}.
Note that we always assume $K(0)=0$. 
Moreover we will assume that that $\sum_{n \in \N} K(n)\le 1$:
the case $\sum_{n \in \N} K(n)< 1$ has to interpreted as the case
of a {\sl terminating} renewal, in the sense that 
$K(\infty):=1-\sum_{n\in \N} K(n)$
is the probability that $\tau_1=+\infty$. Therefore, if $K(\infty)>0$,
the cardinality $\vert \tau \vert$ of the random set $\tau$ is
almost surely finite. The case of $K(\infty)=0$ is instead the case
of a {\sl persistent} renewal, and $\vert \tau\vert=\infty$ 
almost surely. But
 persistent renewals
are of two different kinds: they are {\sl positive} persistent if
$\sum_n n K(n)(= \bE \tau_1) < \infty$, or {\sl null} persistent
if the same quantity diverges. This terminology reflects the fact
that for any {\sl aperiodic} renewal (aperiodicity refers to the fact that $\tau_1$ does not
concentrate on a sublattice of $\N$) the law of large numbers ensures that almost surely
\begin{equation}
\label{eq:RT0}
\lim_{n \to \infty} \frac 1n \big\vert \tau \cap [0, n]\big\vert \, =\, \frac 1{\bE \tau_1}\, \in\, [0, 1],
\end{equation} 
so that if $\bE \tau_1=\infty$ we are facing a {\sl zero density} renewal. 
We stress  that the last statement holds also for terminating renewals
for which $\bE[\tau_1]= \sum_{n \in \N} n K(n) + \infty K(\infty)=\infty$
(note on the way that,
for us,  $\sum_n \ldots$ never includes $n =\infty$).

Very relevant for the analysis of renewal processes is the renewal function 
$n \mapsto \bP (n \in \tau)$, that is the probability that the site $n$ is visited 
by the renewal. We call such a function $K(\cdot)$-{\sl renewal function}
when we want to be more precise. 
The  asymptotic behavior of the renewal function
is captured by the so called Renewal Theorem
(for a  proof see {\sl e.g.} \cite{cf:Asmussen}). This theorem says that
if $\tau$ is an aperiodic renewal (the generalization to the periodic  case is immediate)
we have
\begin{equation}
\label{eq:RT}
\lim_{n \to \infty } \bP (n \in \tau) \, =\, \frac 1{\bE \tau_1} \, \in \, [0, 1].
\end{equation}
Note the link with \eqref{eq:RT0}, but note also that this is little informative if 
$\bE \tau_1=\infty$. The leading asymptotic behavior in such a case 
is summed up in the following statement that calls for the definition
of the Gamma function: $\Gamma (x)=\int_0^\infty t^{x-1}\exp(-t) \dd t$, for $x>0$. 
Recall that in our set-up $\bE[\tau_1]=\infty$ either because $K(\infty)>0$ (terminating renewal), regardless 
of the value of $\ga$,
or because $\ga \le 1$.

\medskip

\begin{proposition}
\label{th:renewal}
Assuming \eqref{eq:K} and \eqref{eq:L} we have:
\begin{enumerate}
\item if $K(\infty)>0$ then
\begin{equation}
\label{eq:renewal1}
\bP (n \in \tau) \stackrel{n \to \infty} \sim \frac{K(n)}{K(\infty)^2}.
\end{equation}
\item If $K(\infty)=0$ and $\ga \in (0,1)$ then
\begin{equation}
\label{eq:renewal2}
\bP (n \in \tau) \stackrel{n \to \infty} \sim 
\left(\frac{\ga \sin (\pi \ga)}{\pi \tc_K} \right)\, n^{\ga -1}.
\end{equation}
\end{enumerate}
\end{proposition}
\medskip

Proposition~\ref{th:renewal}(1) is a classical result detailed for example
in \cite{cf:RegVar} or \cite[\S~A.6]{cf:Book}. Proposition~\ref{th:renewal}(2)
is instead more delicate and while the case $\ga \in (1/2,1)$ is under control
since  \cite{cf:GarsiaLamperti}, the full case is instead a rather recent result
\cite{cf:Doney}. Note however that if full proofs are non-trivial, 
one can rather easily find intuitive arguments 
suggesting the validity of 
Proposition~\ref{th:renewal} \cite[\S~A.6]{cf:Book}.

The asymptotic behavior for the case $\ga=1$ is known too, but this 
case is a bit anomalous and, in this review, we will often skip the results for $\ga =1$ that would make the exposition 
heavier.

\medskip
\begin{rem}
\rm
In full generality, given a renewal $\tau$ one can find a Markov
process $S$  on a state space containing a point $0$ such that
$\tau_1= \inf\{n \in \N :  S_n =0 \}$. Still in full generality, the state space can be chosen
equal to $\N \cup \{0\}$, see \cite[App.~A.5]{cf:Book} for details. Therefore, with this remark in mind, one
could go back to the original definition \eqref{eq:basic} without loss of generality.
\end{rem}
\medskip

\begin{rem}
\rm
\label{rem:SVF}
Everything  we are going to present works assuming simply 
that $K(\cdot)$ is {\sl regularly varying} or, equivalently, that
$L(\cdot)$ is {\sl slowly varying}. Examples of slowly
varying functions include $\log (n) ^c$, $c$ a real number, or
any product of powers of iterated logarithms (see \cite{cf:RegVar}
for full definitions and properties or \cite[\S~A.4]{cf:Book} for
a quick sum-up). Regularly varying functions are a natural set-up
for pinning models also because some natural cases do involve slowly varying functions:
for example the law of $\tau_1$ for the two dimensional simple random walk one
has $K(n) \stackrel{n \to \infty}\sim c/ (n (\log n)^2)$, for an explicit  value of $c>0$ 
\cite{cf:JP}.
\end{rem}

\medskip

In this subsection we have focused on the behavior of the free system: $\gb=h=0$.
The observations made in the caption of Figure~\ref{fig:SRW} do apply 
to the general case too (that is to $(\gb, h) \neq (0,0)$), with, nevertheless, two distinctions:
\smallskip

\begin{enumerate}
\item if $\ga>1$ and if the renewal is persistent, the free renewal is already localized, since by the Renewal Theorem
the contact points have a positive density. We will see that this affects the discussion in the caption 
of  Figure~\ref{fig:SRW} only for what concerns the critical case, but the general picture still holds;
\item if $\tau $ is terminating it is of course harder to localize the process, but in reality it is rather easy to show that 
the model can be mapped to a persistent $\tau$ case, precisely if one sets
$\tilde K(n)= K(n)/(1-K(\infty))$ and if $\tilde \tau$ is the (persistent!)
$\tilde K(\cdot)$-renewal 
\begin{equation}
\label{eq:tildeK}
Z_{N, \go}^\rc \, =\, \bE \left[
\exp\left( \sum_{n=1}^N (\gb \go_n +h +\log (1-K(\infty))) \ind_{n \in \tilde\tau}\right); \, N \in \tilde \tau
\right],
\end{equation}
and the same is true also at the level of the measure $\bP_{N , \go}$ itself. The proof
of such a statement is absolutely elementary and it is detailed for example in \cite[Ch.~1]{cf:Book}:
note that, from a mathematical standpoint, this allows us to restrict ourselves to
persistent  renewals $\tau$. 
\end{enumerate}

\subsection{A gallery of applications}
The localization mechanism captured by class of models 
we have just introduced comes up in  modeling  
a variety of phenomena. Here we just extract some examples
and cite some references.

\subsubsection{Polymers and defect lines} 
The interaction between polymers, chains of elementary units
called monomers, and the surrounding medium or other polymers
is omnipresent in physics, chemistry and biology. We cite for example
\cite{cf:scaling} , but it is of course
impossible to account for the literature in such a direction.
The case we are interested in is the one in which a polymer is fluctuating
in a neutral medium except for a line, or a tube, with which the polymer interacts.
Actually, also cases in which the line is for example a surface or even 
simply a point may be modeled by the type of pinning models we are considering.
The key point is that polymers are often modeled by self-avoiding random walks 
 and a simplified way to impose the self-avoiding condition
is considering directed walks (see Figure~\ref{fig:SRW}). So the polymer 
pinning model becomes precisely the renewal pinning we are considering
(we refer  to \cite{cf:CH,cf:Fisher,cf:GaGr,cf:KNP,cf:Whittington} for
examples of the phenomena that are modeled via directed walk pinning).
Here we just stress that the dimensionality of the problem enters
the renewal pinning only via the exponent $\ga$: for example 
a polymer in three
 dimensions pinned to a line can be modeled by
$\{(n, S_n)\}_{n=0,1, \dots}$, where $S$ is a random walk in two dimensions,
 for which $\ga =0$ (see Remark~\ref{rem:SVF}). The general case
 of a physical space of $d+1$ dimensions leads to $\ga =(d/2)-1$, for
 $d\ge 2$, and of course $\ga =1/2$ if $d=1$.  

\subsubsection{Wetting phenomena} Modeling interfaces in two dimensional media
by random walks has a long history \cite{cf:Abraham} that is somewhat summed up
also in \cite[App.~C]{cf:Book} in which one can find the explanation of why
anisotropic Ising models do reduce in a suitable limit to the renewal pinning model
with $\ga=1/2$.  A particular choice of the boundary conditions leads to
the so called wetting problem \cite{cf:Burkhardt,cf:rough81}, which is just the 
case in which the random walk trajectories that one considers are only
the ones above  (and touching) the axis: with reference to
Figure~\ref{fig:SRW}, to obtain allowed trajectories one has to flip over the negative excursions.
As it is explained in detail in \cite[Ch.~1]{cf:Book}, this problem just corresponds
to renewal pinning with $\ga=1/2$ and $K(\infty)=1/2$ and, at 
the level of contact points,  the process can be mapped  to the case in 
Figure~\ref{fig:SRW} with $h$ replaced by $h-\log 2$ (see \eqref{eq:tildeK}).

\subsubsection{DNA denaturation: the Poland-Scheraga model}
\begin{figure}[htp]
\begin{center}
\leavevmode
\epsfxsize =14.5 cm
\psfragscanon
\psfrag{Loops}[c]{Loops}
\psfrag{A}[l]{\tiny $A$}
\psfrag{G}[l]{\tiny $G$}
\psfrag{C}[l]{\tiny $C$}
\psfrag{T}[l]{\tiny $T$}
\psfrag{l6}[c]{$\tau_{6}-\tau_{5}$}
\psfrag{l9}[c]{$\tau_{9}-\tau_{8}$}
\psfrag{l14}[c]{$\tau_{14}-\tau_{13}$}
\epsfbox{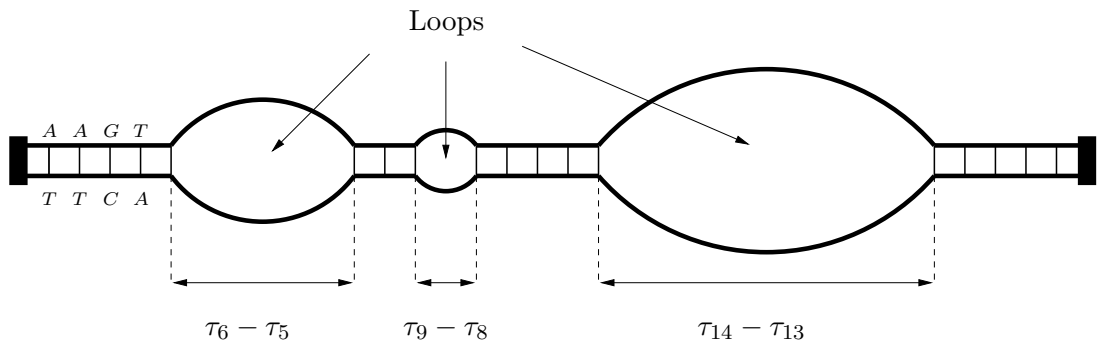}
\end{center}
\caption{\label{fig:DNA} 
The two thick lines are the DNA strands. They may be paired, gaining thus
energetic contributions that depend on whether the base pair is A-T or G-C
(the model is therefore inhomogeneous). There are then sections of unpaired 
bases (the {\sl loops}) to which an entropy is associated. The DNA portion 
in the drawing corresponds to the renewal model trajectory
with $\tau_j-\tau_{j-1}=1$ except  three $\tau$-interarrivals  (so the loops 
correspond to inter-arrival of length $2$ or more).  
}
\end{figure}

Two-stranded DNA has been often modeled by two directed walks 
with pinning potentials, see {\sl e.g.} \cite{cf:Marenduzzo}
and references therein. Since the difference of two independent random
walks is still a random walk, we are back to renewal pinning.
However directed walk models lead to values of $\ga$
that are in contrast with observation, so that a considerable 
amount of work has been put into understanding whether 
(in our language) renewal pinning is a reasonable model and which
$\ga$ should be chosen (see in particular \cite{cf:dna}, but once 
again we refer to \cite{cf:Book} for a more complete bibliography). 
We note that inhomogeneous or disordered modeling is really
required in this context, because the pinning strength does depend
on the type of base pair, see Figure~\ref{fig:DNA}.
The renewal pinning model with inhomogeneous charges has been
and is extensively used for the study of DNA denaturation
\cite{cf:BlosseyCarlon,cf:CH}: appropriate values of $\ga$ are close to $1.15$.

\begin{rem}
\label{rem:nonGauss}
\rm
For DNA denaturation taking $\go_1$ Gaussian is not appropriate.
In this case $\go_1$ should rather be a binary variable.  To be more precise
one should also take into account {\sl stacking } energies, that is energies depend
on blocks of two pairs, and probably  one should also study 
correlated sequences of bases. Sticking to the issue of binary variables versus
Gaussian ones, we take this occasion to stress that much of the mathematical
literature is written for rather general charge distribution (say, with finite exponential
moments of all orders). 
\end{rem}

\section{The homogeneous case}
\label{sec:homogeneous}
The full solution of the non disordered ($\gb=0$), or {\sl homogeneous},  case is crucial and, at the same time,
it is rather elementary once it is phrased in the renewal theory language. 
Such a solution has been repeatedly presented in the physical
literature in several particular instances, by using 
what a probabilist would call {\sl generating function techniques}: 
in particular one can
find a very nice and complete presentation in \cite{cf:Fisher}. 
However the presentation we are going to outline in detail here is different and
 much more direct.

In this section, but also later, with abuse of notation 
we will denote by
$Z_{N, h}^a$ ($a=\rc, \rf$) the partition function $Z_{N, \go}^a$
when $\gb=0$.
Let us start by observing that  we can write
\begin{equation}
\label{eq:basic-hom}
Z_{N, h}^\rc\, =\, \sum_{n=1}^N
\sumtwo{\ell \in \N^n:}{\sum_{j=1}^n \ell_j=N}
\prod_{j=1}^n \exp(h)K(\ell_j).
\end{equation}
Note that if $h=0$ then $Z_{N, h}^\rc=\bP(N \in \tau)$, {\sl i.e.} the partition function is just the 
$K(\cdot)$-renewal function.
The right-hand side of \eqref{eq:basic-hom} is still
a renewal function if $e^hK(\cdot)$ is an inter-arrival law.
And indeed it is if $\sum_{n\in \N} e^h K(n) \le 1$
and in this case $Z_{N,h}^\rc$ is the $e^hK(\cdot)$-renewal function:
its asymptotic behavior is hence given in Theorem~\ref{th:renewal}, but we prefer
to delay such a result since a unified approach holds for every $h$.
In fact if $\sum_{n\in \N} e^h K(n) \ge 1$ we can renormalize
the expression by introducing an exponential correction, going back
to a renewal function (times an exponentially growing factor). Precisely
we call $b(\ge 0)$ the (unique) solution of
\begin{equation}
\label{eq:b}
\sum_{n\in\N} \exp(-bn +h) K(n) \, =\, 1,
\end{equation} 
and we set $K_b(n):= \exp(-bn +h) K(n)$. We have therefore 
defined a function $h \mapsto b(h)$ for $h$ such that
$\sum_{n\in \N} e^h K(n) \ge 1$, that is for $h\ge h_c(0)$, with
\begin{equation}
\label{eq:h_c0}
h_c(0)\, :=\, 
-\log \sum_{n} K(n).
\end{equation}
  For $h< h_c(0)$ we set instead  
$b(h)=0$ and $K_0(n):= \exp(h) K(n)$ (of course
the latter notation is poor since $h$ is not explicit).
With this notations we can write
\begin{equation}
\label{eq:basic-hom1}
Z_{N, h}^\rc\, =\, \exp(bN)
\sum_{n=1}^N
\sumtwo{\ell \in \N^n:}{\sum_{j=1}^n \ell_j=N}
\prod_{j=1}^n K_b(\ell_j)\, =\, \exp(bN) \bP_h( N\in \tau),
\end{equation}
where, under $\bP_h$, $\tau$ is a $K_{b(h)}(\cdot)$-renewal.
By the Renewal Theorem
\begin{equation}
\lim_{N \to \infty}  \bP_h( N\in \tau)\, =\, \frac 1{\bE_h[\tau_1]},
\end{equation}
which is a positive constant if $h>h_c(0)$, but it is zero if $h< h_c(0)$
because the $K_{b(h)}(\cdot)$-renewal is terminating.
 For $h=h_c(0)$
this limit may or may not be zero, but let us postpone this issue 
to Remark~\ref{rem:crit}. Let us focus for now on the fact 
that for $h<h_c(0)$ the Renewal Theorem does not yield
the leading behavior, but thanks to Proposition\ref{th:renewal}(1)
we see that 
\begin{equation}
 \bP_h( N\in \tau) \stackrel{N \to \infty}\sim
 \frac{ K(N)}{\left(1- \exp(h)(1-K(\infty))\right)^2}
 ,
\end{equation}

With these results in our hands we see 
\smallskip
\begin{enumerate}
\item that since $N^{-1}\log  \bP_h( N\in \tau)$ vanishes as $N \to \infty$, we have therefore proven that
\begin{equation}
\label{eq:fe-0}
\tf(0, h)\, :=\, 
\lim_{N\to \infty} \frac 1N \log Z_{N, h}^\rc \, =\, b(h),
\end{equation}
 for every $h$. The quantity $\tf(0, h)$ is usually called {\sl free energy (per unit volume)} and of course we use  such a notation because later there will
 be $\tf(\gb, h)$;
\item that \eqref{eq:basic-hom1} goes well beyond Laplace asymptotics: this is
very relevant and allows us for example to compute the limit of
\begin{equation}
\bP_{N, h}^\rc \left(\tau_1=n_1, \tau_2=n_1+n_2, \ldots, 
\tau_j =n_1 + \dots +n_j\right)\, =\, \prod_{i=1}^j (e^hK(n_i))\frac{Z_{N-n_1-\ldots-n_j,h}^\rc}
{Z_{N,h}^\rc},
\end{equation}
as $N \to \infty$. For example when $h>h_c(0)$ the ratio of partition functions
in the right-hand side converges to $\exp(-(n_1+\ldots+n_j)\tf(0,h))$ and therefore the
all expression converges to $ \prod_{i=1}^j K_{\tf(0,h)}(n_i)$. It is rather easy to see
that the same holds also for $h<h_c(0)$. 
\end{enumerate}
\smallskip

\begin{rem}
\label{rem:crit}\rm
In the above list we have been a bit clumsy about the critical
case $h=h_c(0)$, but in reality what happens at $h=h_c(0)$
is clear too. First of all, $\sum_n K_{\tf(0, h_c(0))}(n)=1$,
so that the associated renewal is persistent. More precisely
$  K_{\tf(0, h_c(0))}(\cdot)=K(\cdot)$ if $\sum_n K(n)=1$,
and $Z_{N , \go}^\rc = \bP (N \in \tau)$,
and otherwise $  K_{\tf(0, h_c(0))}(\cdot)$ is just a multiple of $K(\cdot)$
and $Z_{N , \go}^\rc$ coincides with the $K_{\tf(0, h_c(0))}(\cdot)$-renewal function computed in $N$. Recall now that  the $K_{\tf(0, h_c(0))}(\cdot)$-renewal function
converges to a positive constant if $\ga>1$ and to zero otherwise.
But when it converges to zero there is Proposition~\ref{th:renewal}(2)
that comes to our help so that once again we know the sharp asymptotic behavior of $Z_{N , \go}^\rc$. In particular 
$\tf(0, h_c(0))=0$. 
\end{rem}

\medskip

\begin{figure}[htp]
\begin{center}
\leavevmode
\epsfxsize =14.5 cm
\psfragscanon
\psfrag{0}[c]{$0$}
\psfrag{b}[c]{$h$}
\psfrag{fb}[c]{$\tf(0,h)$}
\psfrag{dfb}[c]{$\partial_h \tf(0,h)$}
\psfrag{hc}[c]{$h_c(0)$}
\epsfbox{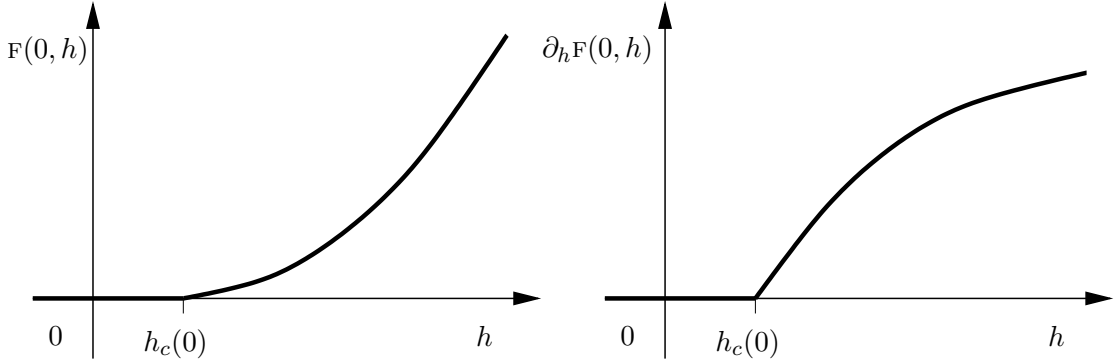}
\end{center}
\caption{\label{fig:fe} 
The function $h\mapsto \tf(0,h)$ is non decreasing, convex and non negative
(convexity follows either from \eqref{eq:b} or \eqref{eq:fe-0}).
 It is therefore equal
to $0$ up to $h_c(0)= \sup \{h:\, \tf(0,h)=0\}$ and after this point it is positive
and strictly increasing. Of course $h=h_c(0)$ is a point of non analiticity 
of the map $h\mapsto \tf(0,h)$: this map is of course analytic on 
$(-\infty, h_c(0))$. It is analytic also on $(h_c(0), \infty)$, by the 
Implicit Function Theorem for analytic functions. The graph of 
$\partial_h \tf(0,h)$ indicates that we are considering the case
$\ga =1/2$: $\tf(0, \cdot)$ is $C^1$ but not $C^2$. For $\ga\in (1/2,1]$ 
the slope of $h\mapsto \partial_h \tf(0,h)$ at $h_c(0)$ is infinite and for $\ga>1$
 a jump discontinuity appears. We stress that $\partial_h \tf(0,h)$ is the contact fraction
 of the system, see Remark~\ref{rem:contact}, and therefore such an observable has a jump at the
 transition for $\ga>1$.}
\end{figure}

All these remarks are telling us in particular that
\eqref{eq:b} is a formula for the free energy, in the sense that
$\tf(0,h)=b$ if there exists a positive solution $b$ to \eqref{eq:b}
and otherwise $\tf(0,h)=0$.
From such a formula one can extract a number of consequences 
that are summed up in the caption of Figure~\ref{fig:fe}. 
In particular the behavior of the free energy near 
criticality is trivial for $h< h_c(0)$, but it is not for $h>h_c(0)$:
let us make explicit 
the behavior of $\tf(0, h)$ as $h\searrow h_c(0)$. 
If $\sum_n n K(n)<\infty$
and if $h_c(0)=0$ (which we may assume without
loss of generality: recall \eqref{eq:tildeK}!)
\begin{equation}
\label{eq:critb1}
\sum_n (1-\exp(-b(h)n))K(n) \, =\, 1-\exp( -h)\stackrel{h\searrow 0}\sim h.
\end{equation}
The asymptotic behavior of the left-hand side
is easily obtained since for every fixed $n $ the limit of $(1-\exp(-b(h)n)/b $ as $b \searrow 0$ is 
 $n$. On the other hand $1-\exp(-x) \le x$ for every $x \ge 0$, so that
 the Dominated Convergence Theorem yields that the left-most side in
 \eqref {eq:critb1} is asymptotically equivalent to
 $b \sum_n nK(n)$, and therefore $b(h) \sim h/\sum_n nK(n)$.
 If instead $\ga \in (0,1)$ formula \eqref{eq:critb1} still holds, but the
 asymptotic behavior of the left-hand side is gotten by 
 Riemann sum approximation:
 \begin{multline}
\label{eq:critb2}
\sum_n \left(1-\exp(-b(h)n)\right) K(n) 
\stackrel{h\searrow 0}\sim
b^{\ga}\tc_K b\sum_n \frac{1-\exp(-b(h)n)}{(bn)^{1+\ga}}
\\
 \sim\, b^{\ga}\tc_K \int_0^\infty 
\frac{1-\exp(-x)}{x^{1+\ga}}\dd x\, =\,  b^{\ga}\tc_K
\frac{\Gamma(1-\ga)}\ga,
\end{multline} 
so that it suffices to invert the asymptotic relation
$ (b(h))^{\ga}\tc_K
(\Gamma(1-\ga)/\ga) \stackrel{h \searrow 0}\sim h$
and this of course gives that $b(h)$ is asymptotically proportional to
$h^{1/\ga}$.

\medskip

The arguments that we have developed directly lead to the following
statement (see \cite[Ch.~2]{cf:Book} for a more complete statement 
and for more details on the proof):

\medskip

\begin{theorem}
\label{th:homogeneous}
The critical behavior of the free energy is given by
\begin{equation}
\tf (0, h_c(0)+ \gd) \stackrel{\gd \searrow 0}\sim
\begin{cases}
c_1 \gd  & \text{ if } \ga >1, \\
c_2 \gd^{1/\ga }  & \text{ if } \ga <1,
\end{cases}
\end{equation}
with 
\begin{equation}
c_1 \, :=\, \frac{1-K(\infty)}{ \sum_{n\in\N} n K(n)} \ \text{ and } \
c_2\, :=\, \left(\frac{\ga (1-K(\infty))}{\tc_K \Gamma (1-\ga)}
\right)^{1/\ga}.
\end{equation}
Moreover in full generality, as $N \to \infty$,  $\bP_{N , h}^\rc$
(that denotes the measure $\bP ^\rc _{N , \go}$ when $\gb=0$)
converges weakly in the product topology of $\R^\N$ to 
a limit measure  $\bP_{h}$. The limit process is a
$K_{\tf(0, h)}(\cdot)$-renewal, namely:
\begin{equation}
\label{eq:bPh}
\bP_{h}\left( \tau_1=\ell_1, \tau_2= \ell_1+\ell_2,
\ldots, \tau_{k}= \ell_1+\ldots+\ell_k \right) \, =\,
\prod_{j=1}^k K_{\tf(0, h)}(\ell_j),
\end{equation} 
where
\begin{equation}
K_{\tf(0, h)}(n )\, =\ \begin{cases}
\exp(h- n\tf(0, h))K(n), & \text{ if } \tf(0, h)>0,
\\
\exp(h)K(n), & \text{ if } \tf(0, h)=0.
\end{cases} 
\end{equation}
Therefore $\tf(0, h)>0$ implies that the limit process
is positive persistent (in fact, the inter-arrival distribution
decays exponentially), while instead if $\tf (0, h)=0$
the limit inter-arrival distribution has power law decay and, if
$h<h_c(0)$, the $K_{\tf(0, h)}(\cdot)$-renewal is terminating.  
\end{theorem}
\medskip

\begin{rem}
\label{rem:free}\rm
Once sharp results for the constrained case are obtained, one can 
deduce sharp results on the free case. This is just based on the 
elementary formula
\begin{equation}
\label{eq:ctof}
Z_{N, h}^\rf\, =\, \sum_{n=0}^{N} Z_{n, h}^\rc \Kbar (N-n),
\end{equation}
that we have written in the case $\sum_{n \in \N} K(n) =1$ and we have introduced
\begin{equation}
\label{eq:Kbar}
\Kbar (N)\, :=\, \sum_{n\in \N:\,  n>N} K(n).
\end{equation}
For example if
 $h>h_c(0)$ from \eqref{eq:basic-hom1}
we have 
\begin{equation}
\label{eq:ctof1}
Z_{N, h}^\rf\, =\, \exp(\tf(0,h) N) \sum_{n=0}^{N} \bP_h(n \in \tau) 
\exp(-\tf(0,h)(N-n))\Kbar (N-n).
\end{equation}
Since  $\exp(-\tf(0,h)(N-n))\Kbar (N-n)$ is bounded below
by $\sum_{j>N-m} \exp(-\tf(0,h)j) K(j)$ and since
$\sum_{n=0}^{N} \bP(n \in \tau) \Kbar (N-n)=1$ for any persistent 
$K(\cdot)$-renewal we obtain that for every $N$
\begin{equation}
\label{eq:lbZf}
Z_{N, h}^\rf\, \ge\, \exp(-h)  \exp(\tf(0,h) N).
\end{equation}
A (rough) bound in the other direction is obtained by 
neglecting $\bP_h(n \in \tau) 
\Kbar (N-n)$ in the right-hand side of \eqref{eq:ctof1}, so that
\begin{equation}
\label{eq:ubZf}
Z_{N, h}^\rf\, \le\, \frac 1{1-\tf (0,h)}  \exp(\tf(0,h) N),
\end{equation}
which holds once again for every $N$.
The sharp asymptotic result is
\begin{multline}
\label{eq:ctof1-1}
Z_{N, h}^\rf\stackrel{N\to\infty} \sim \frac{\exp(\tf(0,h) N)}{\sum_n n K_{\tf(0, h)}(n)}
  \sum_{n=0}^{N} 
\exp(-\tf(0,h) n)\Kbar (n) 
\\
 \sim\, \frac{(1-\exp(-h))\partial_h \tf (0,h)}{1-\exp(-\tf(0,h))} 
 \exp(\tf(0,h) N).
\end{multline}
This type of estimates directly leads to computing the limit behavior
of $\bP_{N, h}^\rf$, see \cite[Ch.~2]{cf:Book} for details.
\end{rem}
\medskip

\begin{rem}
\label{rem:corr} \rm
A key concept in statistical mechanics (and a key concept here)
is the notion of {\sl correlation length}. 
For example a natural correlation length of the system for $h>h_c(0)$
is given by the rate of exponential decay, as $n \to \infty$, of 
$\bP_h( n \in \tau)$ to its limit value $1/\bE_h[\tau_1]$.
One can show \cite{cf:G_EJP}  in particular that  if $h$ is sufficiently
close to $h_c(0)$ then $\bP_h( n \in \tau)-1/\bE_h[\tau_1]>0$ and
\begin{equation}
\lim_{n \to \infty} \frac 1n \log \left(
\bP_h( n \in \tau)-\frac1{\bE_h[\tau_1]}\right) \, =\, - {\tf (0, h)},
\end{equation}
which says that the correlation length coincides with
$1/\tf(0, h)$. Even if one takes a {\sl finite volume viewpoint}, 
$1/\tf(0, h)$ appears as a natural correlation length,
for example
because in \eqref{eq:basic-hom1} one sees that is only when $N$ is of the order of 
$1/\tf(0, h)$ that one starts observing the exponential growth of 
the partition function. One could push these arguments a bit further
and see that if $N$ is {\sl much smaller} than $1/\tf(0, h)$
(of course this has a precise sense only when $h \searrow h_c(0)$)
then $\bP^\rc_{N, h}$ resembles $\bP$, while for $N$ {\sl much larger} than $1/\tf(0, h)$ 
the measure $\bP^\rc_{N, h}$ starts exhibiting localization.
The fact that the inverse of the free energy is the correlation length still
holds also in presence of disorder \cite{cf:GT_Alea,cf:T_jsp,cf:T_ejp},
even if a full understanding of this important issue is still elusive. 
\end{rem}

\medskip

\begin{rem}
\label{rem:contact}
\rm 
The density of contacts, or contact fraction, that is 
 the limit as $N \to \infty$ of
$N^{-1}\bE_h[ \sum_{n=1}^N \ind_{n \in \tau}]$ coincides 
by Theorem~\ref{th:homogeneous}
with
$\lim_{n \to \infty } \bP_{h}^\rc(n \in \tau)=
1/\bE_h \tau_1$. Note moreover that 
$N^{-1}\bE_h[ \sum_{n=1}^N \ind_{n \in \tau}]= N^{-1}\partial_h  \log Z^\rc_{N, h}$,
so that the contact fraction is equal to 
$\partial_h \tf (0, h)$ (except, possibly, at $h=h_c(0)$).
Moreover, by  simple conditioning arguments one easily sees for example that for $h \neq h_c(0)$
\begin{equation}
\lim_{M \to \infty}\lim_{N \to \infty}\max_{n: M\le n \le N-M }
\left\vert \bP_{N, h}^\rc (n\in \tau)- \frac 1{\bE_h \tau_1}\right\vert \, =\, 0, 
\end{equation}
and, by arguing like in Remark~\ref{rem:free}, one directly sees that
he same statement holds for the free case. 
\end{rem}

\section{The disordered case}
\subsection{The quenched free energy}
\label{sec:existence}

An elementary observation that turns out to be really
crucial for us at several instances is that for every $M=0, 1, \ldots, N$
\begin{equation}
\label{eq:superadd}
\log Z_{N, \go}^\rc\, \ge \, \log Z_{M, \go}^\rc +
\log Z_{N-M, \theta^M\go}^\rc.
\end{equation}
It is simply proven by restricting the renewal trajectories, in the 
expression for $Z_{N, \go}^\rc$, 
to the ones that contain the contact site $M$  and by using the renewal property. 
By averaging over the disorder one sees that
$\{ \bbE \log Z_{N, \go}^\rc\}_{N=0,1, \ldots}$ is super-additive and this 
entails \cite{cf:Kingman} the existence of the limit
\begin{equation}
\label{eq:fsuadd1}
\lim_{N \to \infty}
\frac 1N  \bbE \log Z_{N, \go}^\rc \, =:\, \tf (\gb, h),
\end{equation}
and the limit of this sequence coincides with its supremum:
\begin{equation}
\label{eq:fsuadd2}
\tf (\gb, h) \, =\, 
\sup_{N \in \N }
\frac 1N  \bbE \log Z_{N, \go}^\rc .
\end{equation}
The super-additive property \eqref{eq:superadd}
can however be exploited further in order to obtain results 
about the limit of the non averaged sequence
$\{  \log Z_{N, \go}^\rc\}_{N=0,1, \ldots}$, by applying the tools
available for super-additive ergodic sequences, and notably
the celebrated Kingman's Theorem \cite{cf:Kingman}. Alternatively 
one can stick to the super-additive character of 
$\{ \bbE \log Z_{N, \go}^\rc\}_{N=0,1, \ldots}$ and establish
a concentration property  of the non averaged sequence, either by using 
concentration inequalities, {\sl e.g.} \cite{cf:Ledoux,cf:Talagrand}, 
or even by more elementary tools \cite[Ch.~4]{cf:Book}.
In all cases the result that one obtains is 
the existence and the self-averaging character of the free energy 
of pinning systems:

\medskip

\begin{theorem}
\label{th:F}
The sequence $\left\{ N^{-1} \log Z_{N, \go}^\rc\right\}_{N=0,1, \ldots}$ converges to
$\tf(\gb, h)$ both $\bbP(\dd \go)$-almost surely and in the $L^1$ sense.
\end{theorem}

\medskip

\begin{rem}
\label{rem:f}\rm
It is not difficult \cite[Ch.~4]{cf:Book} to show that for every $K(\cdot)$, $\gb$ and $h$ there exists $c>0$
such that
\begin{equation}
\label{eq:f}
Z_{N, \go}^\rc \, \le \, Z_{N, \go}^\rf\, \le c \, N Z_{N, \go}^\rc,
\end{equation}
uniformly in $\go$. We can therefore restate Theorem~\ref{th:F} replacing the superscript $\rc$
with $\rf$.
\end{rem}

\medskip

Another elementary central fact is that
\begin{multline}
\label{eq:basic-el}
Z_{N, \go}^\rc \, \ge \, \bE\left[
\exp\left(
\sum_{n=1}^N ( \gb \go_n +h) \ind_{n \in \tau}
\right); \tau \cap [1,  N] =\{N\}
\right]
= \,\exp(\gb \go_N +h) K(N),
\end{multline}
and therefore
\begin{equation}
\label{eq:0}
\tf (\gb, h ) \, \ge \, 0.
\end{equation}
We now partition the parameter space of the system
into:
\begin{equation}
\label{eq:locdeloc}
\cL \, :=\, \left\{(\gb,h): \, \tf (\gb,h)>0\right\} \ \ \text{ and } \ \ 
\cD \, :=\, \left\{(\gb,h): \, \tf (\gb,h)=0\right\}.
\end{equation}
$\cL$ and $\cD$ stand respectively for $\cL$ocalized and 
$\cD$elocalized regime, a nomenclature that calls for further explanations
(see \S~\ref{sec:paths} just below), but for the moment we just
point out that one of our main aim is to characterize these regions
as precisely as possible. And a substantial help is given by the fact 
that the function
$(\gb, h) \mapsto \tf (\gb, h)$ is convex, as limit of convex functions, 
and it is monotonic 
non decreasing in both variables (monotonicity in $h$ is
immediate, in $\gb$ it is instead a consequence of 
convexity and of the fact that $\partial_h \bbE \log Z_{N, \go}^\rc =0$ for
$\gb=0$). Since by \eqref{eq:0} we 
see that  $\cD$ coincides with $\{(\gb,h): \, \tf (\gb,h)=0\}$ so that
$\cD$ is a convex set. 

\begin{figure}[htp]
\begin{center}
\leavevmode
\epsfxsize =14.5 cm
\psfragscanon
\psfrag{0}[c]{$0$}
\psfrag{b}[c]{$\gb$}
\psfrag{h}[c]{$h$}
\psfrag{hc}[c]{$h_c(0)$}
\psfrag{hcb}[c]{$h_c(\gb)$}
\psfrag{lb}[c]{$\ \ \ h_c^{ann}(\gb):=h_c(0)-\gb^2/2$}
\psfrag{ub}[c]{$\overline{h}(\gb)$}
\psfrag{D}[c]{$\cD$}
\psfrag{L}[c]{$\cL$}
\epsfbox{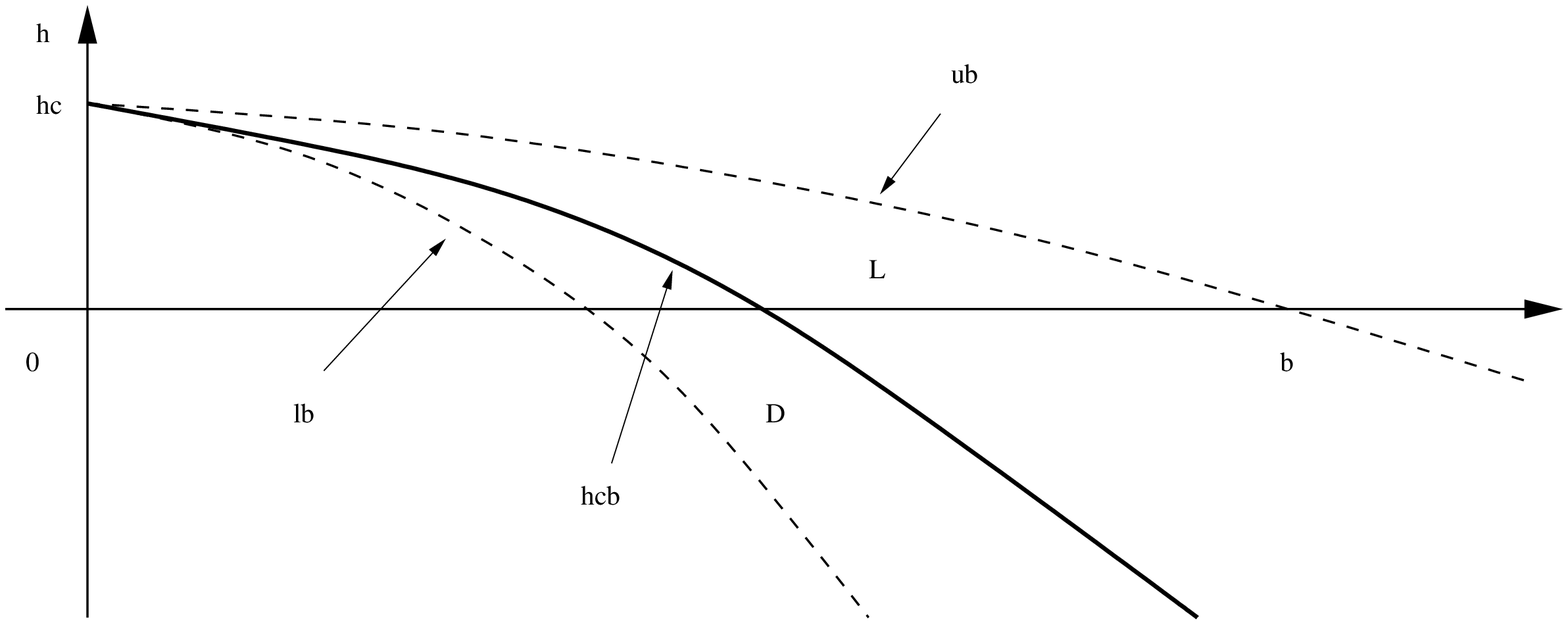}
\end{center}
\caption{\label{fig:window} 
The critical curve $\gb \mapsto h_c(\gb)$ that separates 
$\cD$ and $\cL$ is concave decreasing. This follows from the fact that
$\cD$ is a convex set and from the explicit bounds we have on the critical curve.
The upper bound $\overline{h} (\gb)$ is less explicit then the lower bound, but
we stress that $\overline{h} (\gb)< h_c(0)$ for every $\gb>0$ and this shows that
disorder may induce localization and it never suppresses it. The lower bound 
comes from the standard annealing procedure. Note that the annealed 
partition function $\bbE Z_{N, \go}^a$, $a=\rc, \rf$, is just the homogeneous
partition function with pinning potential $h+\gb^2/2$.}
\end{figure}

One can go beyond:
Jensen inequality ({\sl annealing}) yields
\begin{equation}
\label{eq:annealed0}
\bbE \log Z_{N, \go}^\rc \, \le \,  \log \bbE Z_{N, \go}^\rc \, =\, 
\log \bE \left[\exp\left( (h+\gb^2/2) \sum_{n=1}^N \ind _{n \in \tau}\right); \, N \in \tau
\right] ,
\end{equation}
so that 
\begin{equation}
\label{eq:annealed}
\tf( \gb, h) \, \le \, \tf (0, h+\gb^2/2),
\end{equation}
and if we recall that $\tf(\gb, h)\ge \tf (0, h)$ we directly get
\begin{equation}
\label{eq:sandwich}
  h_c^{ann}(\gb)\, :=\, 
h_c(0)-\frac {\gb^2}2 \, \le \, h_c(\gb) \, \le \,  h_c(0).
\end{equation} 
As a matter of fact the upper bound can be made strict, that is
$h_c(\gb) \, < \,  h_c(0)$ as soon as $\gb>0$, in full generality (the generality here refers to the
choice of $K(\cdot)$,
 \cite{cf:AS}) and in the framework that
we consider here one can show also that, given $K(\cdot)$, for every $\gb_0>0$ 
one can find an explicit constant
$c\in (0, 1/2)$ such that $h_c(\gb) \le  h_c(0) -c\gb^2$  for $\gb \in (0, \gb_0]$ 
\cite[Ch.~5]{cf:Book}. Instead, showing that $h_c(\gb) >  h_c^{ann}(\gb)$ is a more delicate issue
(and it is not true in general!).
These bounds are summed up in Figure~\ref{fig:window}
 and they imply that, since $\cD$ is a convex set, 
then  $h_c(\cdot)$ is concave and, since it is bounded, it is continuous.

Let us sum up the outcome of the arguments we have just outlined:

\medskip
\begin{proposition}
\label{th:hc}
If we set $h_c(\gb)= \inf\{h:\, \tf(\gb, h)>0\}$ then 
$h_c(\gb)= \sup\{h:\, \tf(\gb, h)=0\}$ and
\begin{equation}
\cL \, =\, \left\{ (\gb,h) : \, h> h_c(\gb)\right\} \ \ \text{ and }
\ \ \cD \, =\, \left\{ (\gb,h) : \, h\le h_c(\gb)\right\}
\end{equation}
 Moreover the function
$\gb\mapsto h_c(\gb)$ is concave, decreasing and
\eqref{eq:sandwich} holds for every $\gb$.
\end{proposition}
 \medskip

\begin{rem}
\label{rem:finitevol}\rm
From \eqref{eq:fsuadd2} we actually extract the important observation that
localization can be observed in finite volume, in the sense that
 $(\gb, h)\in\cL$ if and only if 
 there exists $N$ such that $\bbE \log Z_{N, \go}^\rc >0$. 
\end{rem}

\subsection{On path behavior}
\label{sec:paths}

Characterizing localization and delocalization simply by looking
at whether the free energy is positive or zero may look,
from a mathematical standpoint, rather cheap. This is not the case
as one can first see by observing that
\begin{equation}
\partial_h \frac 1N \log Z_{N , \go}^\rc \, =\, \bE_{N,\go}^\rc \left[
\sum_{n=1}^N \ind_{n \in \tau}
\right],
\end {equation}
 which, by exploiting the convexity of the free energy and Theorem~\ref{th:F},
 tells us that $\bbP(\dd \go)$-a.s.
 \begin{equation}
 \label{eq:contact}
 \partial_h \tf (\gb, h)\,  =\,  \lim_{N \to \infty} \bE_{N,\go}^\rc \left[\frac 1N
\sum_{n=1}^N \ind_{n \in \tau}
\right],
 \end{equation}
when $\partial_h \tf (\gb, h)$ exists. 
By convexity, such a derivative exists except possibly at a countable set of points
and in any case \eqref{eq:contact} can be extended to a (standard) suitable statement
also if the derivative does not exist, in terms of right and left derivatives \cite{cf:Book}
(as a matter of fact, in \cite{cf:GT_Alea} it is shown that $\tf (\gb, \cdot)$ is $C^\infty$ except
possibly at $h_c(\gb)$). 
In the end    \eqref{eq:contact}
is telling us that the {\sl contact fraction}, {\sl i.e.} the right-hand side in \eqref{eq:contact}, of our system is zero if $h<h_c(\gb)$ (that is, in the interior of $\cD$) and it is positive
if $h>h_c(\gb)$ (that is, in the whole of $\cL$).
By itself this fully justifies our definition of (de)localization.

However \eqref{eq:contact} is still a poor result and plenty of questions could be
asked about the limit of the sequence $\{ \bP_{N,\go}^a\}_{N=0, 1, \ldots}$,
$a= \rc$ or $a=\rf$, starting with the existence of such a limit. And of course 
the question is: how close one can get to the very sharp description of
the limit measure available for homogeneous systems? 

Work has been done in this direction, but we will not concentrate on this aspect.
We just point out that
\smallskip
\begin{enumerate}
\item The localized phase is, to a certain extent, rather well understood.
In the sense that if $(\gb, h)\in \cL$ than one can show that
the weak limit as $N $ tends to infinity of   the sequence of probability
measures $\{\bP _{N, \go}\}_N$ exists $\bbP(\dd \go)$-a.s. and the limit process
is a point process with a positive density of points \cite{cf:GT_Alea}. One can show also
other estimates going toward the completely clear picture that emerges from the 
homogeneous case. Intriguing differences however do arise, naturally connected
to the existence of exceptional deviations in the sequence of charges. Moreover a number of open
questions still stand (see {\sl e.g.} \cite{cf:GT_irrel}).
\item Progress has been made only recently  on the delocalized phase, at least away of criticality
\cite{cf:GT_PTRF} (see \cite{cf:T_jsp} for some estimates at criticality). Essentially
one now knows that in the delocalized non critical regime the number of contacts
for a system of size $N$ is $O(\log N)$ and such a result has been achieved 
by a subtle argument combining concentration bounds and super-additivity properties
of $\log Z_{N, \go}^\rc$. Such a result
still leaves open  intriguing questions in the direction, for example, of the
precise results proven in \cite{cf:CGZ,cf:IY} in the homogeneous
or weakly inhomogeneous context, see  for example in the bibliographic 
complements at the end of \cite[Ch.~8]{cf:Book}.
\end{enumerate}

\subsection{The role of disorder}
\label{sec:Harris}

The main questions we want to address are:
\smallskip

\begin{enumerate}
\item How does the disorder affect the phase diagram? Namely
can we determine  $h_c(\gb)$, for $\gb >0$, beyond the bounds in
 Figure~\ref{fig:window}?  
\item What can one say about the critical behavior of the free energy?
This amounts to estimating how $\tf (\gb, h)$ vanishes as $h \searrow h_c(\gb)$.
\end{enumerate} 
\smallskip

It is particularly interesting to raise such questions because we know 
$h_c(0)$ and we know the sharp asymptotic behavior of  $\tf (0, h)$ for $h$
close to $h_c(\gb)$ (see Theorem~\ref{th:homogeneous}),
so that in our framework inquiring about the role of the disorder makes 
perfect sense. At this point it is important to underline that such questions
do find partial (non rigorous) answers in the physical literature: the rest 
of this subsection is devoted to explaining what one expects on the basis
of formal expansions, following some renormalization group ideas.
We must say that the arguments that follow are adaptation to the
pinning model context of an argument developed by A. B. Harris \cite{cf:Harris}
in the  context the Ising model with random bond defects. Harris' argument is based on the idea that
the behavior of a system near criticality
should become rather independent of fine details, so in particular one
can replace the system by a coarse grained one without changing substantially the properties.
What one actually tries to do is defining a renormalization transformation, 
like decimation or block summation, 
that, once applied repeatedly at criticality,
transform the system into a limit model. 
 Harris' work aims at determining whether introducing the 
disorder modifies the fixed point of the renormalization transformation:
if the renormalization transformation suppresses the disorder and the limit point
is like in the homogeneous  case, then one says that disorder is 
irrelevant. If instead disorder is enhanced one says that disorder is relevant
and most probably the renormalization transformation flow leads to a fixed point 
which is different from the one obtained in the homogeneous case. It should be noted on one hand that
at the border between relevance and irrelevance 
the renormalization transformation, to first order, neither decreases nor increases the disorder: this is the so called marginal case. On the other hand, Harris argument is just a small disorder expansion 
and as such it does not apply to the whole range of the parameters and, above all,
it does not characterize the limit fixed point when disorder is relevant.

Harris' ideas have been first applied in the pinning model context 
by G. Forgacs, J. M. Luck, Th. M. Nieuwenhuizen and
H. Orland
 \cite{cf:FLNO} and then by B. Derrida, V. Hakim and J. Vannimenus \cite{cf:DHV}
 with predictions that differ somewhat in a sense that we are going to explain just below.

Let us start with an expansion that is freely inspired by  \cite{cf:FLNO}. Without loss of generality
we assume $h_c(0)=0$ (recall \eqref{eq:tildeK}). Moreover
the argument does not feel the boundary condition: we work it out in the free case.
In what follows $\gd:= h+ \gb^2/2 \ge 0$: this change of variable is a natural one
because in particular 
\begin{equation}
\bbE\left[ Z_{N, \go}^\rf\right]\, =\, \bE \left[
\exp\left(\gd \sum_{n=1}^N \ind_{n \in \tau}
\right)
\right]\, =\, 
 Z_{N, \gd}^\rf.
\end{equation}
We
 set $\zeta_n= \exp(\gb \go_n -\gb^2/2)-1$ and let us note that
\begin{equation}
\label{eq:FLNO}
\begin{split}
\bbE \log  \frac {Z^\rf _{N, \go}}{\bbE Z^\rf _{N, \go}}\, &=\, \bbE
\log \bE_{N, \gd}^\rf \left[ \exp \left( \sum_{n=1}^N (\gb \go_n -\gb^2/2)\ind_{n \in \tau}\right)\right]
\\
&=\, 
\bbE \log  \bE_{N, \gd}^\rf \left[ \prod_{n=1}^N (1+\zeta_n \ind_{n \in \tau})\right]
\\
&=\, 
\bbE \log  \left( 1+ \sum_n \zeta_n \bP_{N, \gd}^\rf (n \in \tau) +
\sum_{n_1< n_2} \zeta_{n_1}\zeta_{n_2}\bP_{N, \gd}^\rf (\{n_1, n_2\} \subset \tau) +
\ldots\right).
\end{split}
\end{equation}
Let us now expand the logarithm and let us use the fact that the $\zeta$ random
variables are centered and IID with variance equal to $\exp(\gb^2)-1$ to see that
\begin{equation}
\label{eq:FLNO1}
\bbE \log  \frac {Z^\rf _{N, \go}}{\bbE Z^\rf _{N, \go}}\, =\, - \frac 12
\left(\exp(\gb^2)-1\right)\sum_{n=1}^N \bP_{N, \gd}^\rf (n \in \tau)^2 + \ldots
\end{equation}  
By Remark~\ref{rem:contact}, for $\gd >0$ and as long as $n$ and $N-n$ are large,  
 $ \bP_{N, \gd}^\rf (n \in \tau) $ is close to $\partial_\gd \tf (0, \gd)$ so that
 from \eqref{eq:FLNO} we extract
 \begin{equation}
 \label{eq:FLNO2}
 \tf(\gb, h_c^{ann}(\gb)+\gd )\, =\, 
 \tf(\gb,  \gd-\gb^2/2  ) \, =\, \tf(0, \gd)- \frac 12\left(\exp(\gb^2)-1\right) \left( \partial_\gd \tf (0, \gd)\right)^2 +\ldots
 \end{equation} 
Of course this expansion is only formal and in order  to make it rigorous one has to control the 
rest. Let us note on the way that one can in principle try to compute
all the terms in this expansion, but the issue of controlling the rest is still there and convergence issues
may very well require $\gb $ to be small (note that we are expanding using as small
parameter the variance of $\zeta$, but aiming at capturing the critical behavior, hence $h$
is small too).
All the same,  \eqref{eq:FLNO2} is compatible with $h_c(\gb)=h_c(0)$
if $\tf(0, \gd) $ vanishes much slower than  $( \partial_\gd \tf (0, \gd))^2$ 
as $\gd \searrow 0$ ($\gb$ possibly small, but fixed). But by Remark~\ref{eq:contact} (or directly by
taking the $h$ derivative in \eqref{eq:b}) we see that $\partial _\gd \tf (0, \gd)=
1/\bE _\gd \tau_1$ and by direct computation (similar to \eqref{eq:critb2}) one sees that
$\partial _\gd \tf (0, \gd)\stackrel{\gd \searrow 0}{\sim} (c_2/\ga) \gd^{-1+ 1/\ga}$
for $\ga \in (0,1)$ ($c_2$ is given in Theorem~\ref{th:homogeneous}, but 
the precise value does not play a role here), while the contact fraction 
is bounded away for zero when $\ga >1$ even approaching criticality. 
So \eqref{eq:FLNO2} is compatible with $h_c(\gb)=h_c(0)$ if
\begin{equation}
\label{eq:FLNOcond}
\gd ^{1/\ga} \stackrel{\gd \searrow 0}\gg \gd ^{2(-1+1/\ga)}\ \Longleftrightarrow \ \ga <\frac 12
. 
\end{equation}
This argument therefore suggests that disorder is irrelevant for $\ga <1/2$.

If $\ga >1/2$ the expansion we have performed looks hopeless, but we may argue
that this is just due to the fact that $h_c(\gb)>h_c ^{ann}(\gb)$ and we are
expanding around the {\sl wrong point}. Of course we do know that
$\tf(\gb, h_c(\gb))=0$ and therefore \eqref{eq:FLNO2} suggests 
that for $\gb$ small the shift of the quenched critical point
$\gd_c(\gb):=h_c(\gb)-h_c ^{ann}(\gb)$ is found by equating the two terms in the
rightmost side of \eqref{eq:FLNO2} and this procedure suggests 
$\gd_c(\gb) \approx \gb^{2\ga /(2\ga -1)}$.

\smallskip

A second approach is instead inspired by \cite{cf:DHV}.
If we aim at analyzing whether the annealed system is close to the quenched system
one could sit at the annealed critical point ($h=h_c(0)-\gb^2/2=-\gb^2/2$, {\sl i.e.} $
\gd=0$) and study  the variance
of $Z_{N, \go}^\rf$ (once again, the argument would go through also with constrained
boundary condition).
Divergence of the variance, as $N\to \infty$, would be a sign that quenched and
annealed systems aren't close. Since at $\gd=0$ we have 
$\bbE Z_{N, \go}^\rf=1$ and
\begin{equation}
\text{var}_\bbP \left(Z_{N, \go}^\rf\right)\, =\, 
\bbE\left[ \left(Z_{N, \go}^\rf\right) ^2 -1\right]
\, =\,
\bbE \bE^{\otimes 2}\left[
\exp\left(\sum_n (\gb \go_n -\gb^2/2)(\ind_{n\in \tau}+\ind_{n\in \tau^\prime })
\right) -1
\right]  ,
\end{equation}
with $\tau$ and $\tau^\prime $ independent copies of the same renewal process.
Integrating out the $\go$ variables we obtain
\begin{equation}
\label{eq:DHVvar}
\text{var}_\bbP \left(Z_{N, \go}^\rf\right)\, =\, 
 \bE^{\otimes 2}\left[\exp \left( \gb^2 \sum_{n=1}^N \ind_{n\in \tau\cap \tau^\prime}
\right)-1\right].
\end{equation}
This expression can be evaluated in a sharp way because the random set
$ \tau\cap \tau^\prime$ is still a renewal process and therefore the variance
that we are evaluating is the partition function of a homogeneous pinning model (minus one).
And the first relevant question is whether $ \tau\cap \tau^\prime$ is a terminating
or a persistent renewal. 
The inter-arrival law of $ \tau\cap \tau^\prime$ can be expressed in terms of the 
inter-arrival law of $\tau$ only
in an implicit way, but the renewal function of $ \tau\cap \tau^\prime$ is explicit in terms
of the renewal function of $\tau$:
\begin{equation}
\label{eq:intersect}
\bP^{\otimes 2}\left( n \in  \tau\cap \tau^\prime\right) \, =\, \bP \left( n \in \tau \right)^2,
\end{equation}
and $\tau \cap \tau^\prime$ is terminating (respectively, persistent)
if $ \sum_n \bP \left( n \in \tau \right)^2 < \infty$ (respectively, 
$ \sum_n \bP \left( n \in \tau \right)^2 = \infty$) and, by Proposition~\ref{th:renewal},
we see that
\begin{equation}
\label{eq:DHVcond}
\gamma_2\, :=\, 
 \sum_{n=1}^\infty \bP \left( n \in \tau \right)^2 < \infty \ \
 \Leftrightarrow \ \
 \sum_n \frac 1{n^{2(1-\ga)} } < \infty  
  \ \
 \Leftrightarrow \ \
 \ga <\frac 12
 .
\end{equation}
By the general solution of the homogeneous model, {\sl cf.}
Section~\ref{sec:homogeneous}, we see
that if $\tau \cap \tau^\prime $ is persistent, then for every $\gb >0$
the variance of $Z_{N, \go}^\rf$ grows exponentially,
while if $\tau \cap \tau^\prime $ is terminating then $X:=\vert
\tau \cap \tau^\prime  \vert -1$ is a geometric random variable (this
is just a consequence of the renewal property) of expectation
$\gamma_2$, that is $\bP(X=n)= (\gamma_2/(1+\gamma_2))^n (1/(1+\gamma_2))$,
$n=0,1, \ldots$.
Therefore, as long as $\gb <\gb_0:= \sqrt{\log((1+\gamma_2)/\gamma_2)}$,
 with $p_2:=  (1/(1+\gamma_2))$ we have 
\begin{equation}
\label{eq:3.23}
\lim_{N \to \infty}
\text{var}_\bbP \left(Z_{N, \go}^\rf\right)\, =\, \frac{p_2}{1-(1-p_2)\exp(\gb^2)}-1
\, =\, \gamma_2\gb^2+\ldots
\end{equation}
where the expansion is of $\gb $ small. Therefore if $\tau\cap \tau^\prime $ is terminating
(note that $\tau$ and $\tau^\prime$ are persistent since we are assuming $h_c(0)=0$)
the variance of $Z_{N, \go}^\rf$, at the critical annealed point, stays bounded
and it is small if $\gb$ is small. To complement \eqref{eq:3.23}
note  that
\begin{equation}
\label{eq:3.24}
\sup_N \text{var}_\bbP \left(Z_{N, \go}^\rf\right)\, \le \,
 \frac{p_2}{1-(1-p_2)\exp(\gb^2)}-1
\stackrel{\gb \le \gb_0/2} \le \tilde c \, \gb^2, 
\end{equation}
for some $\tilde c>0$.

\smallskip

Let us sum up the outcome of the arguments we have just outlined:
\begin{enumerate}
\item Both approaches suggest that disorder is irrelevant if 
$\ga<1/2$ (and, as a consequence, relevant if $\ga>1/2$, with
the case $\ga=1/2$ as marginal one), as one can read from 
\eqref{eq:FLNOcond} and \eqref{eq:DHVcond}. Moreover the arguments 
do suggest that the annealed system is very close to the quenched one,
in particular $h_c(\gb)=h_c(0)$, at least for $\gb$ not too large.
This observation may be considered as the Harris criterion prediction for 
pinning models.
\item There is a  difference between \eqref{eq:FLNOcond} and \eqref{eq:DHVcond}
in the case $\ga=1/2$ that we cannot appreciate since we are assuming
\eqref{eq:L}. In the more general framework of Remark~\ref{rem:SVF}
one sees that
 the fact that $L(\cdot)$ diverges at infinity does 
not imply  that $\tau\cap \tau^\prime $ is terminating,
while it is sufficient to conclude that $\tf(0, \gd)$ is much larger
that $(\partial _\gd \tf (0, \gd))^2$ for $\gd $ small.
As a matter of fact, we are dealing
with the marginal case in the renormalization group sense. 
This is a very subtle issue, still unresolved even on  a purely heuristic 
level. We should stress that the steps that we have just presented here
are just a part of the arguments in \cite{cf:FLNO} and \cite{cf:DHV}, in particular
\cite{cf:FLNO} aims at an expansion to all orders and  
\cite{cf:DHV} contains a subtle attempt to study the renormalization
group flow for $\gd$ close to  $0$. Both \cite{cf:FLNO} and \cite{cf:DHV} consider
only the case of $L(\cdot)$  asymptotically constant and, for this
case, their predictions differ. 
\end{enumerate}

\subsection{Relevance and irrelevance of the disorder: the results}
\label{sec:Results}

The heuristic picture outlined in the previous subsection 
has now been made rigorous. A  summary of these rigorous  results is given in the
next three theorems. We recall that $h_c^{ann}(\gb)= h_c(0)-\gb^2/2$
and that the annealed free energy is $\tf (0, h+ \gb^2/2)$, so that
the annealed critical behavior is obtained by looking at $\tf (0, h_c^{ann}(\gb)-(\gb^2/2)
+\gd) =\tf (0, h_c(0)+\gd)$ as $\gd \searrow 0$.

\medskip

\begin{theorem}
\label{th:main<1/2}
Choose $\ga \in(0,1/2)$
and $K(\cdot)$ satisfying \eqref{eq:K} and \eqref{eq:L}.  Then 
 there exists $\gb_0>0$ such that
$h_c (\gb)=h_c^{ann}(\gb)$ for $\gb \le \gb_0$. Moreover, for the same values
of $\gb$ the critical behavior of the quenched free energy coincides 
with the critical behavior of the annealed free energy:
\begin{equation}
\label{eq:exp-behav}
\log \tf (\gb, h_c(\gb)+ \gd) \stackrel{\gd \searrow 0}\sim
\log \tf (0, h_c(0)+ \gd). 
\end{equation} 
\end{theorem}

\medskip

Theorem~\ref{th:main<1/2} has been first proven in 
\cite{cf:Ken} by using a modified second moment method
that we are going to outline in Section~\ref{sec:LB}. It has been then proven 
also in \cite{cf:T_cmp}, by  interpolation  techniques.
Both works contain more detailed results than just  Theorem~\ref{th:main<1/2},
in particular \eqref{eq:exp-behav} has been established by showing that
the stronger statement \eqref{eq:aimLB1} holds.
As a matter of fact in \cite[Th.~2.3]{cf:GT_irrel} it has been proven that
\begin{equation}
\label{eq:limsuplimsup}
\lim_{\beta\searrow0}
\limsup_{\gd\searrow0} \left\vert\frac{\tf(0,\gd)-\tf(\beta,h_c^{ann}(\gb)
+\gd)
}{(\gb^2/2)(\partial_\gd \tf(0,\gd))^2}-1
\right\vert \, =\,0,
\end{equation}
written for $h_c(0)=0$ for sake of compactness. 
Note that \eqref{eq:limsuplimsup} is in agreement with
\eqref{eq:FLNO2} and in fact 
the first step in justifying that expansion.

\medskip

\begin{theorem}
\label{th:main>1/2}
Choose $K(\cdot)$ satisfying \eqref{eq:K} and \eqref{eq:L}.
If $\ga >1/2$ we have $h_c(\gb)> h_c^{ann}(\gb)$. Moreover 
\begin{enumerate}
\item
if  $\ga \in (1/2,1)$ we have that 
for every $\gep>0$ there exists $c_\gep>0$ such that
\begin{equation}
\label{eq:main>1/2}
h_c(\gb)- h_c^{ann}(\gb)\, \ge \, c_\gep \gb^{\frac {2\ga}{2\ga -1}+\gep},
\end{equation}
for every $\gb \le 1$;
\item for every $K(\cdot)$  such
if
 $\ga >1$ there exists $c>0$ such that
\begin{equation}
h_c(\gb)- h_c^{ann}(\gb)\,\ge \,  c\gb^2,
\end{equation}
for every $\gb \le 1$.
\end{enumerate}
\end{theorem}

\medskip

The results in Theorem~\ref{th:main>1/2} are {\sl almost} sharp,
because in \cite{cf:Ken,cf:T_cmp} it is proven that
for every $K(\cdot)$  such
that $\ga \in (1/2,1)$ there exists $C>0$ such that
 \begin{equation}
 \label{eq:UBonshift}
h_c(\gb)- h_c^{ann}(\gb)\, \le \, C \gb^{\frac {2\ga}{2\ga -1}},
\end{equation}
for every $\gb \le 1$. On the other hand the bound in Theorem~\ref{th:main>1/2}(2)
is already optimal (in the same sense) in view of the bounds summed up 
in the caption of Figure~\ref{fig:window}. The result \eqref{eq:main>1/2}
has now been improved to match precisely \eqref{eq:UBonshift}, 
{\sl i.e.} it has been shown in  \cite{cf:AZ} that in \eqref{eq:main>1/2}
one can take $\gep=0$ and $c_0$ is still positive. 

Theorem~\ref{th:main>1/2} has been proven in \cite{cf:DGLT}
and we give an outline of the proof in Section~\ref{sec:UB}.
The method is based on estimating fractional moments of the free energy, 
while 
\eqref{eq:UBonshift} is derived by adapting the techniques yielding Theorem~\ref{th:main<1/2}
(and a sketch of the proof is in Section~\ref{sec:LB}).
In \cite{cf:DGLT} the case $\ga =1$ is not considered, but in
\cite{cf:BS} it is shown that the fractional moment method can be generalized
to establish in particular that $h_c(\gb)>h_c^{ann} (\gb)$ also for $\ga=1$.

\medskip

For what concerns the critical behavior we have the following:
\medskip

\begin{theorem}
\label{th:main_crit}
For every $K(\cdot)$ we have
\begin{equation}
(0\,\le \,  \tf(\gb, h) \, = )\, \tf(\gb, h)-
\tf(\gb, h_c(\gb))\, \le \, \frac{1+\ga}{\gb ^2} \left( h- h_c(\gb)\right)^2,
\end{equation}
for every $h$ (of course the result is  non trivial only for $h>h_c(\gb)$).
\end{theorem}
\medskip

The result in
Theorem~\ref{th:main_crit} has been established in \cite{cf:GT_cmp}
for rather general charge distribution. The proof that we give 
in Section~\ref{sec:smoothing} uses rather heavily the Gaussian 
character of the charges and it is close to the argument sketched in
\cite{cf:GT_prl}.

Let us point out that Theorem~\ref{th:main_crit}, coupled with
Theorem~\ref{th:homogeneous}, shows that the critical behavior of
quenched and annealed systems differ as soon as $\ga >1/2$, in full
agreement with the Harris criterion:
\begin{equation}
\liminf _{\gd \searrow 0}
\frac{
\log \left(\tf(\gb, h_c(\gb)+\gd)-
\tf(\gb, h_c(\gb)))\right)}
{\log \gd}  \stackrel{\ga >1/2} > \, 
\frac{
\log \left(\tf(0, h_c(0)+\gd)-
\tf(0, h_c(0)))\right)}
{\log \gd},
\end{equation} 
since the left-hand side is bounded below by $2$, by Theorem~\ref{th:main_crit}, 
  and the right-hand side
   is equal to $\max(1, 1/\ga)$, by Theorem~\ref{th:homogeneous}.

\medskip

\begin{rem}
\label{rem:smoothing}\rm
Theorem~\ref{th:main_crit} therefore shows that the disorder, for $\ga >1/2$, has 
 a smoothing effect on the transition. The Harris criterion in principle is just suggesting
 that there is no reason to believe that the critical behavior is 
 the same. There is a general belief that disorder smooths the transitions: 
 this is definitely the case for a number of statistical mechanics models to which
 a celebrated result of M. Aizenman and J. Wehr applies \cite{cf:AW} (see also
 \cite{cf:Imry-Ma}). It should
 be however remarked that the Aizenman-Wehr smoothing mechanism does
 not yield smoothing for the pinning model and that the argument leading to
 Theorem~\ref{th:main_crit} is very different from the argument in \cite{cf:AW}
 (for more on this issue see the caption of Figure~\ref{fig:smoothing}). 
\end{rem}  
\medskip

\begin{rem}
\label{rem:smoothing2}\rm
The amount of smoothing proven by Theorem~\ref{th:main_crit} was not fully
expected. In fact in \cite{cf:MG} it is claimed that for $\ga>1$ the transition
is  still  of first order, in disagreement for example with \cite{cf:CH,cf:coluzzi}. 
Needless to say that it would be very interesting to understand what is really
the value of the exponent for $\ga >1/2$ and how it depends on $\ga$.
In \cite{cf:Ken2} it is shown that pinning models based
on exponentially decaying inter-arrival laws  may not
exhibit smoothing.
\end{rem}

\medskip

\begin{rem}
\label{rem:1/2}\rm
The results we have presented do not consider the case $\ga=1/2$.
The results in this case are incomplete and not conclusive under hypothesis
\eqref{eq:L}.
In \cite{cf:Ken,cf:T_cmp} it is shown that under \eqref{eq:L} one
has $h_c(\gb)-h_c^{ann}(\gb) \le c \exp(-1/(c\gb^2))$ for some $c>0$
and $\gb \le 1$. This bound matches the prediction in \cite{cf:DHV}.
It is not known whether $h_c(\gb)-h_c^{ann}(\gb)>0$, leaving open
the possibility for the prediction in \cite{cf:FLNO},
{\sl i.e.} $h_c(\gb)=h_c^{ann}(\gb)$ for $\gb $ small,  to be the right one.
One has to point out however that the approach in \cite{cf:FLNO}
is a development in powers of $\gb$ that cannot capture contributions
beyond all orders. This issue remains open and debated also at a
heuristic level. About the critical behavior, the smoothing result
in Theorem~\ref{th:main_crit} is once again not conclusive
(but it does imply for example that disorder smooths the transition
as soon as $\lim_{n \to \infty} L(n)=0$, if $L(\cdot)$ is chosen as
in Remark~\ref{rem:SVF}). So, under assumption \eqref{eq:L} (which is the
one that arises in the basic example, {\sl cf.} \eqref{eq:tauSRW}), the
issue of whether $\ga=1/2$ is {\sl marginally relevant} or 
{\sl marginally irrelevant} is open. 
\end{rem}

\section{Free energy lower bounds and irrelevant disorder estimates}
\label{sec:LB}

This section is mostly devoted to giving the main ideas of the proof
of Theorem~\ref{th:main<1/2}, but in \S~\ref{sec:LBfor-rel} we will also explain 
why such arguments yield also \eqref{eq:UBonshift}.

\subsection{The case of $\ga<1/2$: the irrelevant disorder regime}
As we have already pointed out, the annealed bound already
yields, in full generality, that $\tf(\gb,h) \le \tf(0, h+ \gb^2/2)$, and hence $h_c(\gb) \ge h_c(0)-\gb^2/2= h_c^{ann}(\gb)$.
In order to pin, for $\ga <1/2$, that $h_c(\gb) \le h_c^{ann}(\gb)$ we need to prove
a lower bound on the free energy showing that
$\tf(\gb,h)>0$ whenever $\tf(0,h+\gb^2/2)>0$. We are actually aiming at
capturing also the critical behavior of the free energy; in fact we are aiming 
at showing  that for every $\gep>0$ there exists $\gb_\gep>0$ such that
for every $\gb \in (0, \gb_\gep)$ 
we have 
\begin{equation}
\label{eq:aimLB1}
\liminf_{h\searrow 0} \frac{\tf(\gb, h)}{\tf(0, h+\gb^2/2)}\,  \ge\,  1-\gep,
\end{equation}
which yields $h_c(\gb) \le h_c^{ann}(\gb)$ and it is a result (sensibly) stronger than
\eqref{eq:exp-behav}.
Note that by what we have seen  on the  expansion
of the free energy \eqref{eq:FLNO2}, or by the rigorous result
\eqref{eq:limsuplimsup},  we cannot aim at proving 
that the quenched free energy coincides with the annealed one.
This actually casts some doubts about the applicability of 
second moment methods. As a matter of fact if we choose $h>-\gb^2/2$ ($h_c(0)=0$), 
as usual with
$\gd=h+\gb^2/2$
we can write 
\begin{equation}
\label{eq:KAvar}
\frac{
\text{var}_\bbP \left(Z_{N, \go}^\rf\right)
}
{\left(\bbE Z_{N, \go}^\rf \right)^2}
\, =\, 
 \bE_\gd^{\otimes 2}\left[\exp \left( \gb^2 \sum_{n=1}^N \ind_{n\in \tau\cap \tau^\prime}
\right)-1\right].
\end{equation}
Note the analogy with \eqref{eq:DHVvar} formula, in which $\gd=0$: this time, since
the underlying measure is $\bP_h$ 
(see the beginning of  Section~\ref{sec:homogeneous}
or Theorem~\ref{th:homogeneous}), $\tau\cap \tau^\prime$ is a positive persistent
renewal and the expression in  \eqref{eq:KAvar} is growing exponentially as $N \to \infty$
for every $\gb>0$ (unlike for the $\gd =0$ case, in which the exponential growth
sets up only for $\gb$ larger than a positive constant $\gb_0$).
As we have pointed out, this was to be expected: can one still
extract from  \eqref{eq:KAvar} some interesting information?
The answer is positive, as shown  by K. Alexander in \cite{cf:Ken}.

The crucial point is not to take the limit in $N$, but rather exploit  
\eqref{eq:KAvar} up to the scale of the correlation length of the annealed system,
which is just a homogeneous system with pinning potential $\gd$ (see 
Remark~\ref{rem:corr}). 
The idea is to establish that the quenched partition function is close to the annealed
one with large probability up to the correlation length scale. Note that,
as pointed out in 
Remark~\ref{rem:corr}, on such a scale
the annealed partition function starts exhibiting exponential growth and 
one feels the size of the free energy (we will actually need to choose the size of the
system, call it 
$N_0$,  to be a large, but finite, 
multiple of the correlation length, the relative error $\gep$ in \eqref{eq:aimLB1}
leaves the room to make such an estimate). Once such an estimate on a system
of length $N_0$
is achieved, it is a matter of chopping the polymer into $N/N_0$
portions: some work at the boundary of these regions is needed and for that we 
refer to  \cite{cf:Ken}, while we focus on explaining why the second moment method
works up to the correlation length scale.

Let us therefore go back to \eqref{eq:KAvar} and let us set $N_0:=q/\tf(0, \gd)$ (and assume
that it is an integer number).
With such a choice, by exploiting the estimates outlined in Remark~\ref{rem:free},
  it is not difficult to see that  for every $K(\cdot)$ and $q>0$
there exists $c_K(q)>0$ such that
\begin{equation}
\label{eq:KAvars1}
\bbE Z_{N_0, \go}^\rf \,=\, \bE \exp \left( \gd \sum_{n=1}^N \ind_{n \in \tau}\right) \, \le \, 
c_K (q).
\end{equation}
This is because, as long as $q$ is finite and $\gd$ tends to zero, the system 
is in the {\sl critical window} (as a matter of fact, in \cite{cf:Julien} 
the limit of $\bbE Z_{N_0, \go}^\rf$ 
as $\gd $ tends to zero, $q$ kept fixed,  is computed and the notion of critical window 
is further elaborated). The constant $c_K(q)$ of course diverges 
as $q\nearrow \infty$.
On the other hand $\bbE Z_{N, \go}^\rf\ge 1$
so that in order to estimate the quantity in \eqref{eq:KAvar}  it suffices to
estimate
\begin{equation}
\label{eq:KAvarest}
 \bE^{\otimes 2}\left[
 \exp\left(\gd \sum_{n=1}^N \left(\ind_{n\in\tau}+
 \ind_{n\in\tau^\prime}\right)\right)
 \left(\exp \left( \gb^2 \sum_{n=1}^N \ind_{n\in \tau\cap \tau^\prime}
\right)-1\right)\right].
\end{equation}
We now use the Cauchy-Schwarz inequality and the fact that
$(\exp(x)-1)^2 \le \exp(2x)-1$ for $x\ge 0$ to bound the
expression in \eqref{eq:KAvarest} (and therefore the expression in
\eqref{eq:KAvar}) by
\begin{equation}
\label{eq:KAvarest2}
 \bE\left[
 \exp\left(2 \gd  \sum_{n=1}^N \ind_{n\in\tau}\right)
 \right]
 \bE\left[
 \left(\exp \left( 2\gb^2 \sum_{n=1}^N \ind_{n\in \tau\cap \tau^\prime}
\right)-1\right)\right]^{1/2}\, =: \, T_1 \cdot T_2.
\end{equation}
But, by \eqref{eq:KAvars1}, $T_1$ is bounded by a constant (which depends on
$q$). On the other hand, $T_2$ has been already estimated in
\eqref{eq:DHVvar}-\eqref{eq:3.24}, and it is $O(\gb)$, thanks to the fact 
that the renewal $\tau\cap \tau^\prime $ is terminating (since $\ga <1/2$).
Therefore, by Chebychev inequality,  for every $\epsilon>0$
\begin{equation}
\bbP\left( Z_{N_0, \go}^\rf \ge (1-\epsilon) \bbE Z_{N_0, \go}^\rf
\right)\, \le \, \frac{C_K(q)}{\epsilon^2} \sqrt{\tilde c}\, \gb,
\end{equation}
where $C_K(q)$ is a constant depending on $K(\cdot)$ and $q$
(it is just the constant $c_K(q)$ of \eqref{eq:KAvars1}  when $\gd$ is replaced by 
$2\gd$) and $\tilde c$ is taken form \eqref{eq:3.24}. 
Since $ \bbE Z_{N_0, \go}^\rf$ is bounded below by $ \exp(-\gd)
\exp(\tf (0, \gd)N_0)= \exp(-\gd+q)$ (see \eqref{eq:lbZf}) we see that 
on the scale of correlation length the quenched partition function 
grows (almost) like the annealed one with large probability if $\gb$ is small enough.

\subsection{Lower bounds on the free energy beyond the irrelevant disorder regime}
\label{sec:LBfor-rel}
The technique for lower bounds on the free energy 
that we have outlined, as well as the 
technique  in \cite{cf:T_cmp,cf:T_school}, 
lead to  upper bounds on $h_c(\gb)$ also in the case $\ga \in [1/2,1)$, see
 \eqref{eq:UBonshift} and Remark~\ref{rem:1/2}. 
 But a look at Theorem~\ref{th:main>1/2}(2) suffices to see that
 the case $\ga > 1$ 
 is somewhat different, because 
it is no longer true that $h_c(\gb) -h_c^{ann}(\gb)= o(\gb^2)$.  
So, in this case,
 the easy bound 
$h_c(\gb) \le h_c(0)$ (which is just a consequence of convexity) is {\sl optimal}
in the sense  that $h_c(\gb)-h_c^{ann}(\gb) \le \gb ^2/2=O(\gb^2)$.  

Let us therefore explain why the second moment method yields also
 \eqref{eq:UBonshift} when  $\ga\in (1/2,1)$. For this we go back
to \eqref{eq:KAvar} and  \eqref{eq:KAvarest} (we are still placing ourselves
on the scale of the correlation length). The term $T_1$ is still bounded
by a constant, that of course depends on $K(\cdot)$ and $q$, just as in the
$\ga <1/2$ case. The term
$T_2$ this time grows exponentially in $N$, because this time $\tau\cap\tau^\prime $
is persistent, and we have to worry about the size of $N$ also for this term.
But let us quickly estimate the growth rate of $T_2$ and for which
values of $N$ we can expect this term to be small for $\gb $ small.
A necessary condition, that with some careful work  one can show also to 
be sufficient, is that the expectation of the term in the exponent
in the exponent is small, namely that ({\sl cf.} Proposition~\ref{th:renewal})
\begin{equation}
\label{eq:KAvar3}
\gb^2 \sum_{n=1}^N \bP(n \in \tau)^2 \, \stackrel{N \to \infty}\sim\,
c_\ga \gb^2 N^{2\ga -1},
\end{equation}
has to be chosen small. However  we still keep $N=N_0= q/\tf (0, \gd)$, that is 
$N$  of the order of $\gd^{-1/\ga}$ (by Theorem~\ref{th:homogeneous}), 
times a constant which is large if $q$
is large. Plugging such a value of $N$ in \eqref{eq:KAvar3}
we see that we are asking  $\gb^2 \gd^{(1-2\ga)/(2\ga)}$ to be small.
Therefore in this regime we expect the second moment method to work,
leading to localization and also to the fact that the 
quenched free energy is fairly close to the annealed one,
if 
\begin{equation}
\gd \, \ge \, c \, \gb^{2\ga/(2\ga -1)},
\end{equation}
with $c $ a small (fixed) constant. But this what is claimed in \eqref{eq:UBonshift}.

\section{Relevant disorder estimates: critical point shift}
\label{sec:UB}

Annealing is the standard procedure to get upper bounds on
disordered partition functions. One can go beyond by partial annealing
procedures, like the {\sl constrained} annealing procedure
\cite{cf:Kuhn}, and this does give some results, see {\sl e.g.} \cite{cf:Ken},
but for pinning models constrained annealing, in the infinite volume limit,  yields nothing beyond
the annealed bound if we are concerned with identifying the critical
point \cite{cf:CG}. There is therefore the need for a different idea.

\subsection{Fractional moment estimates}
\label{sec:UB0}
A  tool that allows to go beyond the annealed bound $h_c(\gb) \ge h^{ann}_c(\gb)$
(we set $h_c(0)=0$ also in this section)
turns out to be
estimating  $A_N:=\bbE[ (Z_{N, \go}^\rc)^\gamma]$
for $\gamma \in (0,1)$ by means of the basic inequality
\begin{equation}
\label{eq:fract}
\left(\sum_j a_j\right)^\gamma \, \le \, \sum_j a_j^\gamma ,  
\end{equation}
that holds whenever $a_j\ge 0$ for every $j$. This 
 has been pointed out by F. L. Toninelli in  \cite{cf:T_AAP}.
Inequality  \eqref{eq:fract} has been exploited also in  other
 contexts, notably  
in \cite{cf:ED,cf:CC}, to get upper bounds 
on the partition function of the directed polymer in random 
environment, and in \cite{cf:AM} to establish
localization of eigenfunctions for random operators, in particular  in the
 Anderson localization context.

 For pinning models  one applies \eqref{eq:fract} 
 to the renewal identity
 \begin{equation}
 Z_{N, \go}^\rc \, =\, \sum_{n=0}^{N-1} Z_{n, \go}^\rc K(N-n) \xi_N, \ \text{ with } \ \xi_N :=
 \exp(\gb \go_N +h),
 \end{equation}
 and, by taking the expectation,  one gets to 
 \begin{equation}
 \label{eq:fhgf}
 A_N \, \le \, \bbE\left[\xi_1^\gamma\right] \sum_{n=0}^{N-1} A_n K(N-n)^\gamma
 \, =\, \sum_{n=1}^N A_{N-n} Q(n),
 \end{equation}
 where $Q_n:=  \bbE[\xi_1^\gamma] K(n)^\gamma$.
 Now the point is that \eqref{eq:fhgf} implies
 \begin{equation}
 A_N \, \le\,  \left(\sum_{n=1}^\infty Q(n)\right)\,  \max_{n=0,1, \ldots,N-1}A_n,
 \end{equation}
 so that, if $\sum_n Q(n) \le 1$ we have 
 $A_N \le A_0=1$ for every $N$. 
 Summing everything up
 \begin{equation}
 \label{eq:gamma0}
  \bbE[\xi_1^\gamma] \sum_{n=1}^\infty 
  K(n)^\gamma \, \le \, 1 \ \Longrightarrow \ \sup_N A_N \le 1. 
 \end{equation} 
 And of course if $A_N$ has sub-exponential growth the free energy is zero
 since
 \begin{equation}
 \frac 1N \bbE \log Z_{N, \go}^\rc \, =\, 
 \frac 1{\ga N} \bbE \log \left(Z_{N, \go}^\rc \right)^\ga \, \le \, 
 \frac 1{\ga N} \log A_N.
 \end{equation}

\medskip
\begin{rem}
\label{rem:polynomial}
\rm 
The discrete convolution inequality \eqref{eq:gamma0}
can actually be exploited more. Observe in fact  that
the solution to the renewal equation 
\begin{equation}
u_0\, =\, 1 \ \text{ and } \ 
u_N\, =\sum_{n=1}^N Q(n) u_{N-n} \ \text{ for } N\, =\,1,2, \ldots,
\end{equation}
 dominates $A_N$.  But if 
$Q(\cdot)$ is a probability distribution (possibly adding $Q(\infty)$), then $u_\cdot$
is the renewal function of the $Q(\cdot)$-renewal and 
if $\sum_n Q(n) <1$ then
$u_N \sim c Q(N)$ for $N\to \infty$ ($c$ is an explicit constant,
see Theorem~\ref{th:renewal}). Therefore
 \begin{equation}
 \label{eq:gamma0-1}
  \bbE\left[\xi_1^\gamma\right] \sum_{n=1}^\infty 
  K(n)^\gamma \, < \, 1 \ \Longrightarrow \ \text{there exists }
  C>0 \text{ such that } A_N \le C K(N)^\gamma. 
 \end{equation} 
\end{rem}

\bigskip

We are now left with verifying for which values of $\beta$ and $h$
we can find $\gamma $ such that 
$ \bbE[\xi_1^\gamma] \sum_{n=1}^\infty 
  K(n)^\gamma \le 1$. Let us consider the case $\sum_n K(n)=1$: in this
  case $\sum_n K(n)^\gamma >1$, so the question is for which
  values of $\gb$ and $h$ the pre-factor $\bbE \xi_1^\gamma $ is sufficiently small.
  This is a straightforward computation ($\gd=h+\gb^2/2$):
 \begin{equation}
 \bbE \xi_1^\gamma  \, =\, \exp\left( -\frac{\gb^2}2  \gamma (1-\gamma) + \gd \gamma\right),
 \end{equation} 
 which is small for $\gb$ sufficiently large, for every fixed value of $\gd$.
 This result is therefore saying that (it may be helpful to keep in mind 
 Figure~\ref{fig:window}):
 \begin{enumerate}
 \item $h_c(\gb)>h_c^{ann}(\gb)$ if $\gb$ is sufficiently large;
 \item the gap between $h_c(\gb)$ and $h_c^{ann}(\gb)$ becomes arbitrarily large as 
 $\gb$ tends to infinity: in fact it is of the order of $\gb^2$.
 \end{enumerate}
 Note that this approach 
 yields very explicit bounds, but
  it does not give results 
 for small values of $\gb$.

\subsection{Iterated fractional moment estimates}
 To go beyond the estimate we have just presented, 
in \cite{cf:DGLT} another renewal identity has been exploited, namely: for every fixed $k$
and every  $N\ge k$
\begin{equation}
  \label{eq:rec2}
  Z^\rc_{N,\omega}\, =\, \sum_{n=k}^N Z^\rc_{N-n,\omega}\sum_{j=0}^{k-1}K(n-j)\,
  {\xi_{N-j}} Z^\rc_{j,\theta^{N-j}\omega}.
\end{equation}
This is simply obtained by decomposing the constrained partition function 
according to the value $N-n$ of the last point of $\tau$ before
 or at $N-k$ ($0\le N-n\le N-k$ in the sum), and to the value $N-j$ of the first point of $\tau$
to the right of $N-k$ (so that $N-k<N-j\le N$).  Of course
$Z^\rc_{j,\theta^{N-j}\omega}$
has the same law as $Z^\rc_{j,\go}$ and the three random variables 
$Z_{N-n,\go}^\rc$, $\xi_{N-j}$ and 
$ Z_{j,\theta^{N-j}\omega}^\rc$
 are independent, if $n\ge k$ and $j<k$.

\begin{figure}[htp]
\begin{center}
\leavevmode
\epsfysize =2.3 cm
\epsfxsize =14.5 cm
\psfragscanon
\psfrag{0}[c]{$0$}
\psfrag{N}[c]{$N$}
\psfrag{N-k}[c]{\small $N-k$}
\psfrag{N-j}[c]{\small $N-j$}
\psfrag{N-n}[c]{\small $N-n$}
\psfrag{Z1}[c]{$Z^\rc_{N-n, \go}$}
\psfrag{Z2}[c]{$K(n-j) \xi_{N-j}$}
\psfrag{Z3}[c]{$Z^\rc_{ N-j,N, \go}$}
\epsfbox{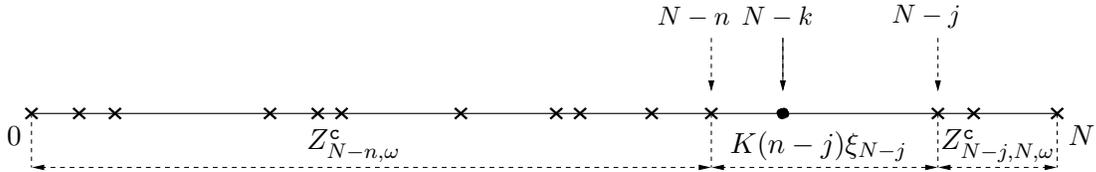}
\end{center}
\caption{\label{fig:chop} 
The renewal identity \eqref{eq:rec2}
is  obtained by fixing a value of $k$ and 
summing over the values of the last contact  before $N-k$ (the large dot in the figure) and the first contact
after $N-k$. The two contacts are respectively $N-n$ and $N-j$ (crosses are contacts
in the figure).}
\end{figure}

\medskip

Let $0<\gamma<1$, set once again  $A_N:=\bbE[(Z^\rc_{N,\omega})^\gamma]$ and use \eqref{eq:fract} and 
 \eqref{eq:rec2}
 to get for $N \ge k$
\begin{equation}
  \label{eq:rec3}
  A_N\le 
\bbE[ \xi_1^\gamma] \sum_{n=k}^N A_{N-n}
\sum_{j=0}^{k-1}K(n-j)^\gamma A_j.
\end{equation}
This is still a renewal type inequality since it can be rewritten as
 \begin{equation}
 \label{eq:k-ren}
 A_N \, \le \, \sum_{n=1}^N A_{N-n} Q_k(n),
 \end{equation}
 with
$Q_k(n):= \bbE[\xi_1^\gamma ] \sum_{j=0}^{k-1}K(n-j)^\gamma A_j$
if $n\ge k$ and $Q_k(n):=0$ for $n<k$. 
In particular if for given
  $\beta$ and $h$ one can find $k\in\N$ and $\gamma\in (0,1)$ such that 
\begin{equation}
  \label{eq:if}
\rho\,:=\, \sum_n Q_k(n)\,=\, \bbE[ \xi_1^\gamma]  \sum_{n=k}^\infty 
\sum_{j=0}^{k-1}K(n-j)^\gamma A_j\, \le\, 1,
\end{equation}
then one directly extracts from \eqref{eq:k-ren} that 
\begin{equation}
  A_N\le \rho \max\{A_0,\ldots,A_{N-k}\},
\end{equation}
for $N \ge k$, which implies that $A_N \le \max\{ A_0, \ldots, A_{k-1}\}$
and hence 
$\tf(\beta,h)=0$.

\medskip

\begin{rem}
\label{rem:polynomial2}
\rm
Like in Remark~\ref{rem:polynomial} one can be sharper
by exploiting the renewal structure in \eqref{eq:k-ren}. 
The difference with Remark~\ref{rem:polynomial} is that in this case $N\ge k$.
In order to put \eqref{eq:k-ren} into a more customary {\sl renewal form} we set
$\tilde A_N:= A_{N}\ind_{N \ge k}$, so that
\begin{equation}
 \label{eq:k-ren2}
 \tilde A_N \, \le \, \sum_{n=1}^{N-(k-1)} \tilde A_{N-n} Q_k(n) + P_k(N), \ \text{ with } \ 
 P_k(N)= \sum_{n=0}^{k-1} A(n) Q_k(N-n), 
 \end{equation}
 and therefore there exists $c>0$ (depending on $k$, $K(\cdot)$ and  
 $\gamma$, besides of course $\gb$ and $h$) such that $P_k(N) \le c Q_k(N)$.
 Let us now consider the standard renewal equation for 
 the $Q_k(n)$-renewal: $u_0=1$ (but of course one can choose an arbitrary 
 $u_0>0$) and 
 \begin{equation}
 \label{eq:k-ren3}
 u_N \, = \, \sum_{n=1}^{N}  u_{N-n} Q_k(n) \stackrel{N \ge k}= 
  \sum_{n=1}^{N-(k-1)} u_{N-n} Q_k(n) + u_0 Q_k(N),
 \end{equation}
 where the first equality holds for $N =1,2, \ldots$ and for the second one 
 we have used  that, since $Q_k(n)=0$ up to $n=k-1$,
 we have $u_1=u_2=\ldots =u_{k-1}=0$. Once again if    $\sum_n Q_k(n)<1$, that is if $\rho<1$,
 we are dealing with a renewal equation of a terminating process and therefore
 $u_N$ behaves asymptotically like (a constant  times) $Q_k(N)$.
 Comparing \eqref{eq:k-ren2} and \eqref{eq:k-ren3} one obtains that there exists
 a constant $C=C(K(\cdot), k, \gamma, h, \gb)$ such that
 \begin{equation}
 A_N \, \le \, C K(N)^\gamma.
 \end{equation}
\end{rem}

\medskip

\subsection{Finite size estimates by shifting}
What the iterated fractional moment has done for us is reducing the problem
of estimating the free energy from above to a finite volume estimate. 
Notice in fact that estimating $\rho$, {\sl cf.} \eqref{eq:if}, amounts to estimating only (a fractional moment of)
$Z_{j, \go}^\rc$, for $j< k$ (note the parallel with Remark~\ref{rem:finitevol}!).
This type of estimates demands a new ingredient, which is more easily explained 
 when $\ga>1$. 
 A preliminary observation is that $\rho$ is bounded above
 by $\gep^{-1}\sum_{j=0}^{k-1} (A_j/(k-j)^{(1+\ga)\gamma -1}$, with $\gep$ a constant that depends
 on $K(\cdot)$ and $\gamma$ so that $\rho \le 1$ is implied if for a given $\gamma$   
 \begin{equation}
 \label{eq:pre-obs}
 \sum_{j=0}^{k-1} \frac{A_j}{(k-j)^{(1+\ga)\gamma -1}} \, \le \, \gep.
 \end{equation}
Note that, unlike the case treated in
 \S~\ref{sec:UB0}, here
  the pre-factor $\bbE \xi_1^\gamma$, and therefore $\gb$
  and $h$,  has only a marginal role: 
 as long as $\gb$ and $h$ are chosen in a compact, which is what we are doing
 since we are focusing on the critical region of the annealed model at small or moderate values of $\gb$, $\gep$ can be chosen independent of the value of $\gb$ and $h$. 
 We shall see that the  expression in \eqref{eq:pre-obs}
can be made small for example by choosing $k$ large. 
 
 Let us start by observing that we know of course (Jensen inequality) that
 for $h=h_c^{ann}(\gb)+\gd$
 \begin{equation}
 \label{eq:annealed2}
 A_j \, \le \, \left(\bbE Z_{j, \go}^\rc\right)^\gamma\, =\, 
 \left(\bE \left[\exp\left(\gd \sum_{n=1}^j \ind_{n\in \tau}\right); \, j \in \tau\right]\right)^\gamma
 \, =\, \exp\left( \gamma \tf(0, \gd) j \right) \bP_\gd(j \in \tau)^\gamma.
 \end{equation}
We are of course interested in $\gd>0$: by the Renewal Theorem
$ \bP_\gd(j \in \tau)$ is bounded below by a positive constant (even if $\gd$ were zero!),
so this term cannot be of much help and we simply bound it above by one.
On the other hand the exponentially growing term stays bounded for 
$j$ up to the correlation length of the annealed system ({\sl cf.}
Remark~\ref{rem:corr}): we therefore choose
$k:=1/\tf(0,\gd)$ (again, assume  that it is in $\N$).
At this point we observe that we can choose $\gamma\in (0,1)$
such that 
\begin{equation}
\label{eq:still>2}
(1+\ga)\gamma\, >\, 2,
\end{equation}
then the expression in 
  \eqref{eq:pre-obs} is bounded for $k$ large, that is $\gd$ small.
  This is not yet what we want, but a more attentive analysis shows that
 one has 
   \begin{equation}
 \label{eq:from-pre-obs1}
 \sum_{j=0}^{k-1-R} \frac{A_j}{(k-j)^{(1+\ga)\gamma -1}} \, \le\, 
 \exp(\gamma) \sum_{j>R} j^{- (1+\ga)\gamma  +1}\, \le
  \, \gep/2,
 \end{equation}
for  any $k\ge R$ and $R$ chosen sufficiently large (depending only
on $\gamma$, $\ga$ and $\gep$). This has been achieved  by using 
  \eqref{eq:annealed2}. We have therefore to show that 
  \begin{equation}
 \label{eq:from-pre-obs2}
 \sum_{j=k-R}^{k-1} \frac{A_j}{(k-j)^{(1+\ga)\gamma -1}} \, \le \, \gep/2.
 \end{equation} 
 For this we set
 \begin{equation}
 \label{eq:toshowA0}
 \hat A _k\, :=\, 
 \limsup_{\gd \searrow 0}
 \max_{j=k-R, \ldots, k-1} A_j.
  \end{equation}
If we are able to show that 
\begin{equation}
 \label{eq:toshowA}
\hat A_k \sum_{i=1}^R i^{-((1+\ga)\gamma -1)} \, \le\,  \gep/3
, 
\end{equation}
then \eqref{eq:pre-obs} would be established (for $\gd$ small
and $k=1/\tf (0, \gd)$. Of course in \eqref{eq:toshowA} one can replace 
$R$ with $\infty$ obtaining thus a more stringent condition (but, in the end,
equivalent, since we are not tracking the consants).
For a proof of  \eqref{eq:toshowA} one has to go beyond
\eqref{eq:annealed2} and in doing so the size of $\gb$ turns out to play a role.

In order to go beyond \eqref{eq:annealed2}
the new idea is a tilting procedure (first proposed in \cite{cf:GLT}), that, given the Gaussian context,
reduces to a shift. The idea is based on the following consequence of 
H\"older inequality
\begin{equation}
\label{eq:Holder}
A_j\, =\, \bbE^\prime  \left[
\left(Z_{j, \go}^\rc\right)^\gamma \frac{\dd \bbP}{\dd\bbP^\prime }(\go)\right]
\, \le \,
\bbE^\prime  \left[
Z_{j, \go}^\rc\right]^\gamma \bbE^\prime  \left[ \left(
 \frac{\dd \bbP}{\dd\bbP^\prime }(\go)\right)^{1/(1-\gamma)} \right]^{1-\gamma},
\end{equation}
where $\bbP^\prime $ is a probability with respect to which $\bbP$ is absolutely
continuous. 
In order to make the 
choice of $\bbP^\prime $ let us fix $\gd =a\gb^2$, $a $ a constant that we are
going to choose along the way:  $\bbP^\prime $ is the law of the sequence
\begin{equation}
\go_1- \sqrt{a \gb^2}, \go_2- \sqrt{a \gb^2}, \ldots , \go_{k}- \sqrt{a \gb^2}, 
\go_{k+1}, \go_{k+2},\ldots
\end{equation}
which is a sequence of independent (non identically distributed) variables.
One then readily computes 
\begin{equation}
\bbE^\prime  \left[ \left(
 \frac{\dd \bbP}{\dd\bbP^\prime }(\go)\right)^{1/(1-\gamma)} \right]^{1-\gamma}\, =\, 
\exp\left( \frac{\gamma}{1-\gamma} a \gb^2 k\right), 
\end{equation}
but $a \gb^2 k =\gd/\tf(0,\gd )$ and 
 the ratio $\gd/\tf(0,\gd )$ tends to a positive constant
as $\gd \searrow 0$ since $\ga >1$, {\sl cf. } Theorem~\ref{th:homogeneous}. 
Let us now turn our attention to $\bbE^\prime Z_{j, \go}^\rc$
which, for $j\le k$,  coincides with $\bbE^\prime Z_{j, \go- \sqrt{a \gb^2}}^\rc$.
But this is just the partition function of a homogeneous model with negative
pinning potential if we choose $a$ small, 
namely (for conciseness we look only at the case
 $j=k$)
\begin{equation}
\label{eq:j=k}
\begin{split}
\bbE^\prime Z_{k, \go}^\rc \, &=\, \bE\left[
\exp\left( -\gb^2 (\sqrt{a}-a) \sum_{n=1}^k \ind_{n \in \tau}\right); \, k\in \tau \right]
\\
&=\, 
\bE\left[
\exp\left( - \left(\frac1{\sqrt{a}}-1\right) \left(\frac{\gd}{\tf(0,\gd )}\right)
 \frac 1k\sum_{n=1}^k \ind_{n \in \tau}\right); \, k\in \tau \right].
\end{split}
\end{equation}
But $\lim_{k \to \infty} (1/k)\sum_{n=1}^k \ind_{n \in \tau} = 1/\bE[\tau_1]$ $\bP$-a.s.
and this readily implies that $\bbE^\prime Z_{k, \go}^\rc$ is bounded by a constant that
can be chosen arbitrarily small, provided one chooses $a$ sufficiently small
(so $a^{-1/2}-1$ is large). 
The argument easily extends to $j$ between $k-R$ and $k$ so that
\eqref{eq:toshowA} is proven since $\hat A_k$
can be chosen arbitrarily small for $a$ sufficiently small
and every $\gd \le \gd_0$ (for some $\gd_0>0$).
This concludes the argument for the case $\ga>1$, that is 
Theorem~\ref{th:main>1/2}(2).

\smallskip

The case $\ga \in (1/2,1)$ (Theorem~\ref{th:main>1/2}(1)) is conceptually not very different: the main difference lies 
in the fact that it is no longer sufficient to show that
$A_j$ is small for $j$ close to $k$, one has actually to extract some decay in $j$.
But the fact that $A_j$ does decay with $j$, at least if $j \le 1/\tf(0, \gd)$,
is already rather evident from \eqref{eq:annealed2} from the fact that 
the term $\bP_\gd( j\in \tau)$ is, at least till $j <k=1/\tf(0, \gd)$, close to
$\bP( j\in \tau)$ which behaves for $j$ large as $j^{-(1-\ga)}$ (times a constant,
{\sl cf.} Theorem~\ref{th:renewal}). The argument is however somewhat technical
and we refer to \cite{cf:DGLT} for details.

\section{Relevant disorder estimates: the critical exponent}
\label{sec:smoothing}

\begin{figure}[htp]
\begin{center}
\leavevmode
\epsfysize =2.3 cm
\epsfxsize =14.5 cm
\psfragscanon
\psfrag{0}[c]{\small $0$}
\psfrag{l}[c]{\small $\ell$}
\psfrag{2l}[c]{\small $2\ell$}
\psfrag{9l}[c]{\small $9\ell$}
\psfrag{G1}[c]{\small $G_1\ell$}
\psfrag{G2}[c]{\small $G_2\ell$}
\psfrag{G3}[c]{\small $G_3\ell$}
\psfrag{G4}[c]{\small $G_4\ell$}
\epsfbox{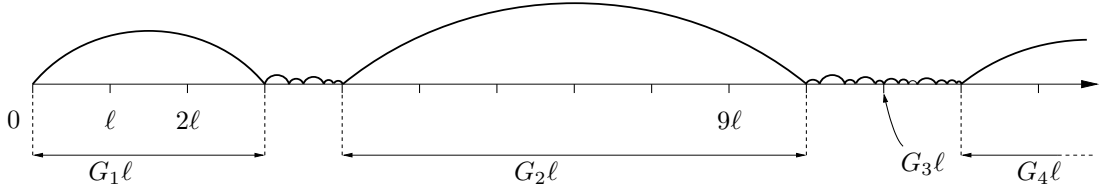}
\end{center}
\caption{\label{fig:smoothing} 
The rare stretch strategy is implemented by looking at blocks of $\go$ variables of size $\ell$.
To block $j$ is associated a Bernoulli random variable $Y_j$: such a random variable is a function of the $\go$
variables in the block and it determines whether a system of the size of that block, with
constrained boundary conditions and precisely with the $\go$ variables of the block, has a 
sufficiently large (in fact, an atypically large) partition function. The precise definition of 
$Y_j$ is in \eqref{eq:Yj}. In the figure $Y_4=Y_{11}=Y_{12}=1$, while the other $Y$
variables are zero. The gaps between the success blocks are parametrized as $G_n \ell$,
so that $\{G_n\}_n$ are IID Geometric random variables: $\bbP(G_1=k)= (1-p)^{k}p$
($k=0,1,2, \ldots$ and $p=p(\ell)$, given in the text).  The lower bound is then achieved by restricting the partition function to the 
$\tau$ trajectories that visit only success blocks and  visit the first and the last
point of the block (also if two success blocks are contiguous: this
is the case of the second and third success blocks in the figure). 
Such a strategy is profoundly different from that employed in \cite{cf:AW},
that is based on the effect of typical fluctuations (on the Central Limit Theorem scale)
and competition with boundary effects. Our strategy is instead a Large Deviation
strategy and, in a sense, it  exploits the {\sl flexibility} of the polymer to target
rare regions (the boundary conditions play no role).  
}
\end{figure}

The argument leading to Theorem~\ref{th:main_crit} is based on the {\sl rare stretch} strategy
sketched in Figure~\ref{fig:smoothing}.
It is based on a one-step  coarse graining of the environment
on the scale $\ell\in \N$, $1\ll \ell \ll N$. Actually one should think of 
$\ell $ as very large but finite. We assume $N/\ell \in \N$ and
we look at the sequence of IID random variables defined as
\begin{equation}
\label{eq:Yj}
Y_j\, :=\, \ind_{E_j}, \ \text{ with } \ E_j=\left\{ \go: \, 
\log Z_{\ell, \theta^{(j-1)\ell}\go}^\rc \ge a\tf(\gb, h+\gd )\right\},
\end{equation}
where $\gd>0$, $a\in (0,1)$ (eventually $a \nearrow 1$) and  
 $j=1,2, \ldots$, but of course only the $j$'s up to $N/\ell$ are relevant to us.
Note that the $Y$ variables are  Bernoulli  random variables of parameter
$p(\ell):= \bbP( E_1)$ and, since $a<1$,  $p(\ell)$ is small when $\ell$ is large,
by the very definition of the free energy and its self-averaging property
({\sl cf.} Theorem~\ref{th:F}). 
One can actually show rather easily that
\begin{equation}
\label{eq:shiftsmooth}
\liminf_{\ell \to \infty}\frac 1\ell \log p(\ell) \, \ge \,- \frac {\gd^2}{2 \gb^2}.  
\end{equation}
We give a proof of this inequality below, but the intuitive reason is that
the probability of observing $\sum_{i=1}^\ell \go_i \approx \ell \gd /\gb$
 behaves like $\exp(-\ell \gd^2 /2\gb^2)$ for $\ell$ large (this is the standard
 Cramer Large Deviation result). When such a Large Deviation event occurs,
 the environment in the $\ell$-block will look like the original $\go$ variables
 translated of $\gd/\gb$, that is $\gb \go_i+h$ looks like
 $ \gb \go_i+h+\gd$. And in that block the logarithm of the partition function
 will hence be close to $\ell \tf(\gb, h+\gd)$. Shifting the mean is of course
 only one possible strategy to make the event $E_1$ typical and hence 
 such an argument yields only a lower bound on $p(\ell)$.

 We now make a lower bound on $Z_{N , \go}^\rf$ by considering 
 only the $\tau$-trajectories that visit all and only the $\ell$
 blocks for which $Y_j(\go)=1$ (see Figure~\ref{fig:smoothing}).
 The renewal property leads to a rather explicit lower bound, namely (with
 the notation of the figure)
 \begin{equation}
\log Z_{N , \go}^\rf \, \ge \, \sumtwo{j\le N/\ell:}{Y_j=1} \log Z^\rc_{\ell,
\theta^{(j-1)\ell}\go}\, +\, \sum_{n=1}^{\cN_Y(\go)} \log K(G_n\ell) +
O(\log N),
 \end{equation}
where the $O(\log N) $ term comes from the last excursion. Let us now divide
by $N$ and take the limit, keeping into account that, by definition of the
$Y$ variables, we have a lower bound on the partition functions 
$ \log Z^\rc_{\ell,
\theta^{(j-1)\ell}\go}$ that
appear in the right-hand side. We therefore obtain
\begin{equation}
\label{eq:jabv}
\begin{split}
\tf(\gb, h)\, &\ge \, a \ell \tf(\gb, h+\gd) \lim_{N \to \infty} \frac {\cN_Y(\go)}N\, +\,
\limsup_{N \to \infty} \left(\frac {\cN_Y(\go)}N \frac 1{\cN_Y(\go)}\sum_{n=1}^{\cN_Y(\go)} \log K(G_n\ell) \right)
\\
&=\, a p(\ell) \tf(\gb, h+\gd) + \frac{p(\ell)}\ell 
\limsup_{N \to \infty} \frac 1{\cN_Y(\go)}\sum_{n=1}^{\cN_Y(\go)} \log K(G_n\ell),
\end{split}
\end{equation} 
where in the last step we have used the strong law of large numbers (the limits
are in the $\bbP(\dd \go)$--a.s. sense) to estimate the leading behavior of 
the number of successes in an array of $N/ \ell$ Bernoulli variables of 
parameter $p(\ell)$. Moreover, to be precise, \eqref{eq:jabv} holds if we set $K(0)=1$
(which we do only here).
The (superior) limit that is left in the expression is also easily 
evaluated by using the strong law of large numbers after having observed
that, since $\log K(x) \stackrel{x\to \infty}\sim -(1+\ga )\log x$,
 when $G_n=1, 2, \ldots$ we have $\log K(G_n\ell)\ge -a^{-1} (1+\ga) (\log G_n+\log \ell)$
for $\ell$ sufficiently large (uniformly in the value of $G_n$: note that $1/a$ is once again just
a number larger than $1$). 
The net outcome is therefore
\begin{equation}
\tf(\gb, h)\, \ge \, a p(\ell) \tf(\gb, h+\gd ) - \frac{p(\ell)}\ell a^{-1} (1+\ga) \left(\bbE[\log G_1; G_1>0]+\bbP (G_1>0) \log \ell\right),
\end{equation}
Since $G_1$ is a geometric variable of parameter $p(\ell)$ we directly compute
$\bbE[\log G_1; G_1>0]= (1+o_\ell(1))\log (1/p(\ell))$ which, by \eqref{eq:shiftsmooth},
is bounded above by $a^{-1}\ell \gd^2/(2\gb^2)$ ($\ell $ large: once again
$a^{-1}$ is just used as an arbitrary constant larger than one). Therefore
\begin{equation}
\label{eq:ineqs}
\tf(\gb, h)\, \ge \,  p(\ell) \left[a\tf(\gb, h+\gd ) - a^{-2}\frac {\gd^2 (1+\ga)}{2\gb^2}+ o_\ell(1)\right],
\end{equation}
where the term $o_\ell(1)$ is $-c(\log \ell) /\ell$ ($c>0$)
and this bound holds, given  $a\in (0,1)$,
for every $\ell$ larger than some $\ell_0$.

Now we set $h=h_c(\gb)$ in \eqref{eq:ineqs}, so that the left-hand side is zero and therefore
\begin{equation}
a\tf(\gb, h_c(\gb)+\gd ) - a^{-2}\frac {\gd^2 (1+\ga)}{2\gb^2}+ o_\ell(1) \, \le \, 0,
\end{equation}
for every $ \ell>\ell_0$, so that $\tf(\gb, h_c(\gb)+\gd) \le  a^{-3}({\gd^2 (1+\ga)}/{(2\gb^2)})$,
and since $a\in (0,1)$ is arbitrary we can let $a\nearrow 1$ and we are done.

\medskip

For completeness we give a proof of \eqref{eq:shiftsmooth}. We call
$\tilde \bbP_\ell$ the law of the sequence of random variables
\begin{equation}
\go_1+\gd/\gb, \go_2+\gd/\gb, \ldots , \go_\ell+\gd/\gb, \go_{\ell+1},  \go_{\ell+2}, \ldots
\end{equation}
Note that $\bbP _\ell(E_1)$ tends to one as $\ell $ becomes large, simply 
by definition of free energy and 
because
the law of $\{\gb\go_n+h\}_{n=1, \ldots , \ell}$, when $\go$ is distributed according to
 $\tilde\bbP_\ell$,
 coincides with the law of $\{\gb\go_n+h+\gd\}_{n=1, \ldots , \ell}$, when 
 $\go$ is distributed according to
 $\bbP$.
We compute the relative entropy
\begin{equation}
\cH \left( \tilde\bbP_\ell \big \vert \bbP \right) \,:=\,
\tilde \bbE_\ell \left[\log \frac{\dd \tilde \bbP_\ell}{\dd\bbP}(\go)
\right] \, =\, \frac{\gd^2}{2\gb^2}\ell, 
\end{equation}
and by a standard entropy inequality (see {\sl e.g.} \cite[\S~A.2]{cf:Book})
\begin{equation}
\log \left( \frac{\bbP(E_1)}{\tilde \bbP_\ell (E_1)}\right)\, \ge \, 
-\frac 1{\tilde \bbP_\ell (E_1)} \left( \cH \left( \tilde\bbP_\ell \big \vert \bbP \right) + \frac 1e
\right)\, =\, 
-\frac 1{\tilde \bbP_\ell (E_1)} \left( \frac{\gd^2}{2\gb^2}\ell + \frac 1e
\right),
\end{equation}
and this yields \eqref{eq:shiftsmooth}.

\section*{Acknowledgments}
The author is very grateful to the organizers of the Summer School on Spin Glasses
for the invitation to lecture on disordered pinning models. 
He also acknowledges the support of ANR, grant POLINTBIO.


\begin{thebibliography}{99}

\bibitem{cf:Abraham}
D. B. Abraham,
\emph{Surface Structures and Phase Transitions, Exact Results},
in {\sl Phase Transitions and Critical Phenomena} {\bf 10}, 
Academic Press, London (UK)  (1986), 1-74.

\bibitem{cf:AM}
M. Aizenman and S. Molchanov, \emph{Localization at large disorder and at extreme energies: an elementary
derivation}, Commun. Math. Phys. {\bf 157} (1993), 245-278.

\bibitem{cf:AW}
M.~Aizenman and  J.~Wehr, \emph{
Rounding effects of quenched randomness on first-order phase transitions},
 Commun. Math. Phys. {\bf 130} (1990),  489-528. 


\bibitem{cf:Ken} K.~S.~Alexander, \emph{The effect of disorder on
    polymer depinning transitions}, Commun. Math. Phys. {\bf 279} (2008),
     117-146.
     
     \bibitem{cf:Ken2} K.~S.~Alexander, \emph{
Ivy on the ceiling:  first-order polymer depinning transitions with quenched disorder},  Markov Proc.  Rel. Fields {\bf 13} (2007), 663-680.      
      
  \bibitem{cf:AS} K. S. Alexander and V. Sidoravicius, {\sl Pinning of polymers and interfaces by random potentials}, Ann. Appl. Probab. {\bf 16} (2006), 636-669.
  
  \bibitem{cf:AZ} K.~S.~Alexander and N.~Zygouras,
\emph{Quenched and annealed critical points in polymer pinning models},
arXiv:0805.1708 [math.PR]

\bibitem{cf:Asmussen} 
 S.~Asmussen, \emph{Applied probability and queues}, Second Edition,
Applications of Mathematics {\bf 51}, Springer-Verlag, New York (2003).




\bibitem{cf:RegVar} N. H. Bingham, C. M. Goldie and J. L. Teugels,
  \emph{Regular variation}, Cambridge University Press, Cambridge,
  1987.

\bibitem {cf:BS} 
M.~Birkner and  R.~Sun,
\emph{Annealed vs quenched critical points for a random walk pinning model}, 
	arXiv:0807.2752v1 [math.PR]

\bibitem{cf:BlosseyCarlon}
R.~Blossey and E.~Carlon, 
\emph{
Reparametrizing the loop entropy weights: effect on DNA melting curves},
  Phys. Rev. E{\bf 68} (2003), 061911  (8 pages).

\bibitem{cf:BCT} E. Bolthausen, F.  Caravenna and  B. de Tili\`ere, {\sl
The quenched critical point of a diluted disordered polymer model},
arXiv:0711.0141v1 [math.PR]


\bibitem{cf:Burkhardt}
T.~W.~Burkhardt, \emph{Localization-Delocalization 
transition in a solid-on-solid model with a pinning potential},
  { J. Phys. A: Math. Gen.} {\bf  14} (1981), L63-L68.

\bibitem{cf:CC} A.~Camanes and P. Carmona, \emph{Directed polymers, critical temperature
and uniform integrability}, preprint (2007).

\bibitem{cf:CG}
F.~Caravenna and G.~Giacomin,
\emph{On  constrained annealed bounds 
for pinning and wetting models},
{ Elect. Comm. Probab.}  {\bf 10} (2005), 179-189.

\bibitem{cf:CGZ}
F. Caravenna, G. Giacomin and L. Zambotti,
\emph{A renewal theory approach to periodic copolymers with adsorption}, 
Ann. Appl. Probab. {\bf 17} (2007), 1362-1398.

 \bibitem{cf:CH}
 D.~Cule and  T.~Hwa, \emph{
  Denaturation of Heterogeneous DNA}, 
   Phys. Rev. Lett. {\bf 79} (1997), 2375-2378.


\bibitem{cf:coluzzi} B. Coluzzi and E. Yeramian, {\sl Numerical
evidence for relevance of disorder in a Poland-Scheraga DNA
denaturation model with self-avoidance: scaling behavior of average
quantities}, Eur. Phys. Journal B {\bf 56} (2007), 349-365.


\bibitem{cf:DGLT}
B. Derrida,
G. Giacomin, H. Lacoin and F. L. Toninelli, {\it
Fractional moment bounds and disorder relevance for pinning models},
arXiv:0712.2515 [math.PR]


\bibitem{cf:DHV} B. Derrida, V. Hakim and J. Vannimenus,
  \emph{Effect of disorder on two-dimensional wetting}, J. Statist.
  Phys. {\bf 66} (1992), 1189-1213.

\bibitem{cf:Doney} R. A. Doney, \emph{One-sided local large deviation and 
renewal theorems in the case of infinite mean}, Probab. Theory Rel. Fields
{\bf 107} (1997), 451-465.

\bibitem{cf:ED}
M. R. Evans and  B. Derrida, \emph{Improved bounds for the transition temperature of directed polymers in a
finite-dimensional random medium}, J. Statist. Phys. {\bf 69} (1992), 427-437.

\bibitem{cf:Feller1}  
W.~Feller, \textit{An introduction to probability theory and its applications}, Vol. I,  
 Third edition, John Wiley \& Sons, Inc.,   
New York--London--Sydney, 1968. 

\bibitem{cf:Fisher}
M.~E.~Fisher, \emph{ Walks, walls, wetting, and melting},
J. Statist. Phys. {\bf 34} (1984), 667-729.

\bibitem{cf:FLNO} G. Forgacs, J. M. Luck, Th. M. Nieuwenhuizen and
H. Orland, \emph{Wetting of a disordered substrate: exact critical
behavior in two dimensions}, Phys. Rev. Lett. {\bf 57} (1986),
2184-2187.

\bibitem{cf:GaGr}  
S.~Galluccio and  and  R.~Graber, 
\emph{Depinning transition of a directed polymer by a periodic potential: a $d$-dimensional
solution},  Phys. Rev. E {\bf 53} (1996), R5584-R5587.



\bibitem{cf:GarsiaLamperti} A. Garsia and J. Lamperti, \emph{
A discrete renewal theorem with infinite mean},
Comment. Math. Helv. {\bf 37} (1963), 221-234.

\bibitem{cf:scaling}
P. G. de Gennes,  \emph{
Scaling concepts in polymer physics},
Cornell University Press, Ithaca, NY (1979).

\bibitem{cf:Book} G. Giacomin, \emph{Random Polymer Models},
Imperial College Press, World Scientific (2007).

\bibitem{cf:G_EJP}
G. Giacomin,  
\emph{Renewal convergence rates and correlation decay for homogeneous pinning models}, 
Elect. J. Probab. {\bf 13} (2008), 513-529. 

\bibitem{cf:GLT} G. Giacomin, H. Lacoin and F. L. Toninelli, \emph{
    Hierarchical pinning models, quadratic maps and quenched
    disorder}, arXiv:0711.4649 [math.PR]

\bibitem{cf:GT_PTRF} G.~Giacomin and F.~L.~Toninelli, \emph{Estimates on
    path delocalization for copolymers at selective interfaces},
  Probab. Theor. Rel. Fields {\bf 133} (2005), 464-482.

\bibitem{cf:GT_prl}
G.~Giacomin and  F.~L.~Toninelli, 
\emph{Smoothing of Depinning Transitions for Directed Polymers with Quenched Disorder},
Phys. Rev. Lett. {\bf  96} (2006), 060702.

\bibitem{cf:GT_cmp} G.~Giacomin and  F.~L.~Toninelli, \emph{
Smoothing effect of quenched disorder on polymer depinning transitions},
Commun. Math. Phys. {\bf 266} (2006), 1-16.

\bibitem{cf:GT_Alea}
G.~Giacomin and F.~L.~Toninelli, 
\emph{The localized phase of disordered copolymers with adsorption}, 
ALEA {\bf 1} (2006),149-180.

\bibitem{cf:GT_irrel} G. Giacomin and F.~L.~Toninelli, {\sl On the
    irrelevant disorder regime of pinning models}, 
    arXiv: 0707.3340v1 [math.PR]

  \bibitem{cf:Harris} A.~B.~Harris, {\sl Effect of Random Defects on
      the Critical Behaviour of Ising Models}, J. Phys. C {\bf 7}
    (1974), 1671-1692.
    
 \bibitem{cf:Imry-Ma} 
 Y.~Imry  and  S.--K.~Ma, 
\emph{Random-field instability of the ordered state of continuous symmetry},
 Phys. Rev. Lett. {\bf 35} (1975), 1399-1401.


\bibitem{cf:IY}
 Y.~Isozaki  and  N.~Yoshida, \emph{
Weakly pinned random walk on the wall:
pathwise descriptions of the phase Transition},
 Stoch. Proc. Appl.  {\bf 96}(2001),   261-284.

\bibitem{cf:JP}
N.~C.~Jain and W.~E.~Pruitt, \emph{
The range of random walk}, in {\sl Proceedings of the Sixth Berkeley Symposium on Mathematical Statistics and Probability} (Univ. California, Berkeley, Calif., 1970/1971), Vol. III: Probability theory, 
31-50. Univ. California Press, Berkeley, Calif., 1972.   
 
\bibitem{cf:dna} Y. Kafri, D. Mukamel and L. Peliti, \emph{Why is the
DNA denaturation transition first order?}, Phys. Rev.
Lett. {\bf 85}
(2000), 4988-4991.

\bibitem{cf:KNP}
Y.~Kafri, D.~R.~Nelson and A.~Polkovnikov,
\emph{
Unzipping flux lines from extended defects in type-II superconductors},
 Europhys. Lett. {\bf 73} (2006), 253-259.


\bibitem{cf:Kingman} 
J. F. C.
 Kingman, \emph{
 Subadditive Ergodic Theory}, { Ann. Probab.} {\bf 1} (1973), 882-909.	

\bibitem{cf:Kuhn}
R.~K\"uhn, \emph{
 Equilibrium ensemble approach to disordered systems I: 
 general theory, exact results}, {  Z. Phys. B} {\bf 100} (1996), 
 231-242.

 
 \bibitem{cf:Ledoux}
M.~Ledoux,
\emph{The concentration of measure phenomenon}, 
{ Mathematical Surveys and Monographs}, {\bf 89}, American
Mathematical Society, Providence, RI (2001).
	
	\bibitem{cf:rough81}
	J.~M.~J.~van Leeuwen and H.~J.~Hilhorst, 
\emph{Pinning of rough interface by an external potential},
  Phys. A {\bf 107} (1981), 319-329.	
  	   				   
\bibitem{cf:Marenduzzo}
D.~Marenduzzo, A.~Trovato and A.~Maritan,  \emph{
Phase diagram of force-induced DNA unzipping in exactly solvable models},
{ Phys. Rev. E} {\bf 64} (2001), 031901 (12 pages).				   
				   
				   
\bibitem{cf:MG} C. Monthus and T. Garel, \emph{
 Distribution of pseudo-critical temperatures and
  lack of self-averaging in disordered Poland-Scheraga 
  models with different loop exponents},
 Eur. Phys. J. B {\bf  48} (2005), 393-403.

\bibitem{cf:Petrelis}
N.~P\'etr\'elis, \emph{
Polymer Pinning at an Interface}, Stoch. Proc. Appl. {\bf 116} (2006), 1600-1621.

\bibitem{cf:Julien} J. Sohier,
\emph{ 
Finite size scaling for homogeneous pinning models},
arXiv:0802.1040 [math.PR]


\bibitem{cf:Talagrand}
M.~Talagrand, 
\emph{A new look at independence}, { Ann. Probab.} {\bf 24} (1996), 1-34.
 
\bibitem{cf:T_jsp}
F.~L.~Toninelli, \emph{Critical properties and finite-size estimates for 
the depinning transition of directed random polymers}, J. Statist. Phys. {\bf 126} (2007), 1025-1044. 
 
 \bibitem{cf:T_ejp}
F.~L.~Toninelli, \emph{Correlation lengths for random polymer models and for some renewal sequences}, Electron. J. Probab.  {\bf 12} (2007), 613-636. 
 
\bibitem{cf:T_cmp}  F.~L.~Toninelli,
\emph{A replica-coupling approach to disordered pinning models},
Commun. Math. Phys. {\bf 280} (2008), 389-401.

\bibitem{cf:T_AAP} F.~L.~Toninelli, \emph{Disordered pinning models and
    copolymers: beyond annealed bounds},  Ann. Appl.
  Probab. {\bf 18} (2008), 1569-1587.
  
  
  
  
  \bibitem{cf:T_school}
F.~L.~Toninelli, \emph{Localization transition in disordered pinning models. Effect of randomness on the critical properties}, Lecture Notes from the $5^{\rm th}$ 
 Prague Summer School on 
 Mathematical Statistical Mechanics, September  2006,   arXiv:0703912
[math.PR].

\bibitem{cf:Whittington}
S.~G.~Whittington,\emph{
A directed-walk model of copolymer adsorption},
{ J. Phys. A: Math. Gen.} {\bf 31} (1998), 8797-8803.


\end{thebibliography}
\end{document}